\documentclass[onecolumn]{mn2e}
\usepackage{graphicx}
\def\pmb#1{\mbox{\boldmath$#1$}}
\def\gtsim {>\kern-1.2em\lower1.1ex\hbox{$\sim$}}
\def\ltsim {<\kern-1.2em\lower1.1ex\hbox{$\sim$}}
\begin{document}
\title[Inertial modes of neutron stars with the superfluid core]{Inertial modes of neutron 
stars with the superfluid core}
\author[S. Yoshida and U. Lee]{Shijun Yoshida$^1$\thanks{E-mail: yoshida@fisica.ist.utl.pt} and 
Umin Lee$^2$\thanks{E-mail: lee@astr.tohoku.ac.jp} \\
$^1$Centro Multidisciplinar de Astrof\'{\i}sica -- CENTRA,
           Departamento de F\'{\i}sica, Instituto Superior T\'ecnico, \\
           Av. Rovisco Pais 1, 1049-001 Lisboa, Portugal \\
$^2$Astronomical Institute, Graduate School of Science,
           Tohoku University, Sendai 980-8578, Japan
}
\date{Typeset \today ; Received / Accepted}
\maketitle
\begin{abstract}
We investigate the modal properties of inertial modes of rotating neutron stars 
with the core filled with neutron and proton superfluids, taking account of entrainment 
effects between the superfluids.
In this paper, the entrainment effects are modeled by introducing 
a parameter $\eta$ so that no entrainment state is realized at $\eta=0$.
We find that inertial modes of rotating neutron 
stars with the superfluid core are split into two families, which we call ordinary fluid
inertial modes ($i^o$-mode) and superfluid inertial modes ($i^s$-mode). The two 
superfluids in the core counter-move for the $i^s$-modes. For the $i^o$-modes,  
$\kappa_0=\lim_{\Omega\rightarrow 0}\omega/\Omega$ is only weakly dependent 
on the entrainment parameter $\eta$,
where $\Omega$ and $\omega$ are the angular frequency of rotation and 
the oscillation frequency observed in the corotating frame of the star, respectively. 
For the $i^s$-modes, on the other hand, $|\kappa_0|$ almost linearly increases as 
$\eta$ increases.
Avoided crossings as functions of $\eta$ are therefore quite common 
between $i^o$- and $i^s$-modes.
We find that some of the $i^s$-modes that are unstable against the gravitational radiation 
reaction at $\eta=0$ become stable when $\eta$ is larger than $\eta_{crit}$, the value of
which depends on the mode. 
Since the radiation driven instability associated 
with the current multipole radiation is quite weak for the inertial modes and
the mutual friction damping in the superfluid core is strong, 
the instability caused by the inertial modes will be easily suppressed
unless the entrainment parameter $\eta$ is extremely small and the mutual friction damping
is sufficiently weak.
\end{abstract} 

\section{Introduction}

Since Andersson (1998) and Friedman \& Morsink (1998) found that the $r$-modes 
of a rotating neutron star are 
unstable against gravitational radiation reaction even for infinitesimally small rotation speeds, 
many papers on the $r$-mode instability have appeared.
As a result of these investigations,
the importance of the instability in astrophysics has been
pointed out mainly in the following two ways, that is,
the instability is strong enough to limits both
the spin of newly born neutron stars and that of old accreting 
neutron stars, and neutron stars in which the $r$-mode oscillation is excited
will be a potential source of gravitational waves that are detectable by the LIGO
and other detectors (Lindblom, Owen, \& Morsink 1998; Andersson, Kokkotas, \& Schutz 
1999; Owen et al. 1998; Bildsten 1998; Andersson, Kokkotas, \& Stergioulas 1999).
For a recent review, see, e.g., Andersson \& Kokkotas (2001). 

It has long been suggested that neutrons in the inner crust and
neutrons and protons in the core of neutron stars are
in superfluid states when the interior temperatures 
cool down below $T\sim 10^9{\rm K}$ (e.g., Shapiro \& Teukolsky 1983).
Since the interior temperature of neutron stars is believed to cool down quickly
by emitting neutrinos (e.g., Baym \& Pethick 1979),
we may expect
that many observable neutron stars, except newly born ones, 
are likely to have a core that contains superfluids.
Although one of the important ingredients that affect the oscillations of neutron stars is 
superfluidity in the interior, it was rather recently that
Epstein (1988) pointed out the importance of superfluidity 
for the oscillations, suggesting
the existence of a new family of acoustic modes 
associated with superfluidity by means of the local dispersion relations derived
for the case of non-rotating neutron stars.
Later on, Lindblom \& Mendell (1994) studied the $f$-modes of rotating
neutron stars that have a core filled with neutron and proton superfluids, taking account of 
entrainment effects between the superfluids, and 
showed that the 
ordinary fluid-like $f$- modes are not strongly affected by the superfluidity in the core.
In the same paper, Lindblom \& Mendell (1994) suggested 
the existence of superfluid-like acoustic modes, which they called $s_0$-modes, but
they could not find the $s_0$-modes numerically.
First numerical examples of the superfluid-like acoustic modes of 
neutron stars with the superfluid core were obtained by Lee (1995), who also
showed that $g$-modes are not propagating in the core filled with superfluids.
Andersson \& Comer (2001) discussed the dynamics of 
superfluid neutron star cores 
examining local dispersion relations in detail, and 
they confirmed that $g$-modes are excluded from the superfluid core (Lee 1995).
Applying their argument to the $r$-modes, 
they suggested that the $r$-modes of a superfluid neutron star core
are split into two distinct families, that is,
ordinary fluid-like $r$-modes and superfluid-like $r$-modes.
Most recently, Prix \& Rieutord (2002) have computed in great detail
acoustic modes of non-rotating neutron stars composed of 
neutron and proton superfluids
for a wide range of the entrainment parameter.
Although their results are basically consistent with
the results obtained by Lee (1995) and Andersson \& Comer (2001),
they have also shown that the assumption that
$|\delta\pmb{v}_p-\delta\pmb{v}_n|\sim0$ for the ordinary fluid-like acoustic modes and
$|\rho_p\delta\pmb{v}_p+\rho_n\delta\pmb{v}_n|\sim0$ for the superfluid-like acoustic modes
is not necessarily helpful to distinguish the two 
families, where $\delta\pmb{v}_p$ and $\delta\pmb{v}_n$
are the Eulerian velocity perturbations of the proton and neutron superfluids, 
the mass densities of which are denoted as $\rho_p$ and $\rho_n$. 
We also note that 
Sedrakian \& Wasserman (2000) examined 
rotational effects on the fundamental modes in uniform density superfluid stars,
applying the so-called tensor virial technique to normal modes.
As for relativistic non-rotating superfluid stars, Comer, 
Langlois, \& Lin (1999) calculated superfluid-like acoustic modes and showed that 
the basic features of the superfluid modes are quiet similar to those of the
corresponding Newtonian superfluid modes. More recently, Andersson, Comer \& Langlois 
(2002) have found avoided crossings between the superfluid- 
and the ordinary fluid-like acoustic modes in a relativistic superfluid
neutron star as the entrainment is varied.

To give an answer to the question whether the $r$-mode instability 
will survive the damping due to 
mutual friction in the superfluid core of cold neutron stars,
Lindblom \& Mendell (2000) computed ordinary fluid-like $r$-modes, 
taking account of the entrainment effects in the core.
They showed that the mutual friction damping in the core
is not effective to suppress the instability except for a few domains 
of the entrainment parameter $\eta$.
Recently, Lee \& Yoshida (2002) have studied the $r$-modes 
in neutron stars with the superfluid core 
with a different method, confirming most of the results
by Lindblom \& Mendell (2000).
Lee \& Yoshida (2002) have also found that, as suggested by Andersson \& Comer (2001), 
the $r$-modes are split into ordinary fluid-like
$r$-modes and superfluid-like $r$-modes, and that the instability 
caused by the superfluid-like $r$-modes
is extremely weak and easily damped by dissipation processes in the interior.

The purpose of this paper is to study modal properties of inertial modes
in rotating neutron stars containing superfluids in the core. 
As shown by Lockitch \& 
Friedman (1999), the existence of inertial modes in the lowest order of angular frequency $\Omega$
of rotation depends on the structure of the so-called zero-frequency subspace of 
eigensolutions to linearized hydrodynamic equations. 
Naively speaking, we may say that, for inertial modes to exist in the limit of $\Omega\rightarrow 0$,
$g$-modes of a non-rotating star with $\Omega=0$ need to be
degenerate to the zero-frequency subspace.
Investigating the zero-frequency subspace of pulsation equations for the superfluid 
core, Andersson \& Comer (2001) showed the existence of the zero-frequency $g$-modes, 
and hence the existence of inertial modes (see, also, Comer 2002). 
In this paper, we 
calculate global inertial mode oscillations in neutron stars with a core filled 
with neutron and proton superfluids, taking account of the entrainment effects 
between the two superfluids. 
For the dynamics of superfluids in the core, 
we employ a two-fluid formalism similar to that given in Lindblom \& Mendel (1994). 
To obtain the inertial modes in the limit of $\Omega\rightarrow 0$, we generalize 
the method devised by Lockitch \& Friedman (1999) to
stars composed of a superfluid core and a normal fluid envelope.  
In \S 2 we present the basic equations employed in 
this paper for the dynamics of superfluids in the core, and in \S 3 dissipation 
processes considered in this paper are described. \S 4 gives numerical results,
and \S 5 and \S 6 are for discussions and conclusions.

\section{oscillation Equations}

We derive basic equations governing superfluid motions in the neutron star core in 
the Newtonian dynamics, assuming uniform rotation of the star. The core is assumed 
to be filled with neutron and proton superfluids and a normal fluid of electron.
We also assume perfect charge neutrality between the protons and electrons because
the plasma frequency is much higher than the oscillation frequencies considered in
this paper (see, e.g., Mendell 1991a). Since the transition temperatures $T_c\sim 10^9$ 
K to neutron and proton superfluids are much higher than the interior temperatures of 
old neutron stars (see, e.g., Epstein 1988), we assume that all the neutrons and 
protons in the core are in superfluid states and the excited normal fluid components 
of the fluids can be ignored.

In this study, we assume that the neutron and proton superfluids and the electron 
normal fluid in the equilibrium state are in the same rotational motion with the 
angular velocity $\Omega$ around the axis of rotation.
The velocity of the fluid $v^a$ in the equilibrium state is therefore given by 
$v^a=\Omega\,\varphi^a$, where $\varphi^a$ is the rotational Killing vector.  
In a perturbed state, however, the neutron and the proton superfluids can move 
differently from each other in the core, obeying their own governing equations.
In order to describe the perturbations in the superfluid core, thus, we can choose  
the perturbed neutron density $\delta \rho_n$, the perturbed proton density 
$\delta \rho_p$, the perturbed velocity of the neutron $\delta v^a_n$, and 
the perturbed velocity of the proton $\delta v^a_p$ as dynamical variables. 
Since the equilibrium star is axisymmetric about the rotation axis, we assume that 
the time and azimuthal dependence of the perturbations is given by
$\exp(i\sigma t+i m\varphi)$ with $\sigma$ being the oscillation frequency observed
in an inertial frame, and $m$ is an integer representing the azimuthal wave-number.
The basic perturbation equations employed in this paper for the neutron and proton 
superfluids in rotating neutron stars are essentially the same as those given in 
Mendell (1991a) and Lindblom \& Mendell (1994) and are given as follows: 
\begin{equation}
i \omega\delta\rho_n+\nabla_a\,(\rho_n\delta V_n^a)=0, 
\label{con-n}
\end{equation}
\begin{equation}
i \omega\delta\rho_p+\nabla_a\,(\rho_p\delta V_p^a)=0,
\label{con-p}
\end{equation}
\begin{eqnarray}
i \omega\delta v_n^a+2\Omega\,\delta v_n^b\nabla_b\varphi^a+
\nabla^a(\delta\mu_n+\delta\Phi)-2\Omega\,\frac{\rho_{np}}{\rho_n}\, 
(\delta v_p^b-\delta v_n^b)\,\nabla^a\varphi_b = 0 \, ,
\label{eul-n}
\end{eqnarray}
\begin{eqnarray}
&&i \omega\left(\delta v_p^a+\frac{m_e}{m_p}\,\delta v_e^a\right)+
2\Omega\,\left(\delta v_p^b+\frac{m_e}{m_p}\,\delta v_e^b\right)\,\nabla_b\varphi^a
\nonumber \\
&&+ \nabla^a\left(\delta\mu_p+\frac{m_e}{m_p}\,\delta\mu_e+\left(1+\frac{m_e}{m_p}
\right)\,\delta\Phi\right)-2\Omega\,\frac{\rho_{np}}{\rho_n}\,
(\delta v_n^b-\delta v_p^b)\,\nabla^a\varphi_b = 0 \, ,
\label{eul-p}
\end{eqnarray}
where $\omega$ denotes the corotating frequency, defined by $\omega=\sigma+m\Omega$, 
$\nabla_a$ means the covariant derivative in flat three 
dimensional space, and $\delta Q$ stands for the Eulerian change in the physical 
quantity $Q$. Here, $m_p$ and $m_e$ are the proton and the electron masses.
In the mass conservation equations (\ref{con-n}) and (\ref{con-p}), $\rho_n\delta V_n^a$ 
and $\rho_p\delta V_p^a$ mean the mass current vector of the neutron and of the proton, 
respectively, and can be expanded in terms of the perturbations of the superfluid 
velocities as follows: 
\begin{equation}
\rho_n\delta V_n^a=\rho_{nn}\,\delta v_n^a+\rho_{np}\,\delta v_p^a \, , 
\end{equation}
\begin{equation}
\rho_p\delta V_p^a=\rho_{pn}\,\delta v_n^a+\rho_{pp}\,\delta v_p^a \, , 
\end{equation}
where the relationships $\rho_{n}=\rho_{nn}+\rho_{np}$, 
$\rho_{p}=\rho_{pn}+\rho_{pp}$, and $\rho_{np}=\rho_{pn}$ must be satisfied due to 
the Galilean invariance of equations (\ref{con-n}) and (\ref{con-p}). 
The perturbed superfluid velocities are then given in terms of $\delta V_n^a$ 
and $\delta V_p^a$ as
\begin{equation}
\delta v_n^a={\rho_{11}\over\rho_n}\,\delta V_n^a
+{\rho_{12}\over\rho_n}\,\delta V_p^a \, ,
\end{equation}
\begin{equation}
\delta v_p^a={\rho_{21}\over\rho_p}\,\delta V_n^a
+{\rho_{22}\over\rho_p}\,\delta V_p^a \, ,
\end{equation}
where
\begin{equation}
\rho_{11}={\rho_{pp}\rho_n^2\over\tilde{\rho}^2}, \quad
\rho_{22}={\rho_{nn}\rho_p^2\over\tilde\rho^2}, \quad
\rho_{12}=\rho_{21}=-{\rho_{np}\rho_n\rho_p\over\tilde\rho^2},
\end{equation}
and
\begin{equation}
\tilde\rho^2=\rho_{nn}\rho_{pp}-\rho_{np}\rho_{pn}.
\end{equation}
Note that $\rho_{11}+\rho_{12}=\rho_n$ and $\rho_{22}+\rho_{21}=\rho_p$.
In Euler equations (\ref{eul-n}) and (\ref{eul-p}), $\delta v_e^a$ and $\delta\Phi$ are 
the perturbed velocity of the electron and the perturbed gravitational potential, 
respectively, and $\delta\mu_n$, $\delta\mu_p$, and $\delta\mu_e$ stand for the 
perturbed chemical potentials per unit mass for the neutron, proton and electron, 
respectively.  Because of the assumption of the perfect charge neutrality in the proton 
and electron plasma, we may have 
\begin{equation}
\delta V_p^a=\delta v_e^a, \quad {\rm and} \quad \delta \rho_p/m_p=\delta \rho_e/m_e,
\end{equation}
where $\rho_e$ is the electron density. The perturbed gravitational potential is 
determined by the Poisson equation 
\begin{equation}
\nabla^2\delta\Phi=4\pi G\delta\rho,
\end{equation}
where $\rho=\rho_n+\rho_p+\rho_e$ and $G$ is the gravitational constant.

Using a variant of Gibbs-Duhem relation, the pressure perturbation is given by
\begin{equation}
\delta p=\rho_n\delta\mu_n+\rho_p\delta\tilde\mu \, , 
\end{equation}
where $\delta\tilde\mu=\delta\mu_p+\frac{m_e}{m_p}\,\delta\mu_e$. 
Here, we have ignored the entropy carried by the electron normal fluid.
This treatment is justified for a star whose temperature is low enough.   
Note also that the superfluids carry no entropy.

To obtain a relation between the densities $\rho_n$ and $\rho_p$ and the chemical potentials
$\mu_n$ and $\tilde\mu$, we begin with writing the energy density $e$ as
\begin{equation}
e=e\left(\rho_n,\rho_p\right), 
\end{equation}
with which the chemical potentials are defined as
\begin{equation}
\mu_n\left(\rho_n,\rho_p\right)=\left(\partial e/\partial \rho_n\right)_{\rho_p}, \quad
\mu_p\left(\rho_n,\rho_p\right)=\left(\partial e/\partial \rho_p\right)_{\rho_n}.
\end{equation}
If the chemical potential of the electron is given by
\begin{equation}
\mu_e(\rho_e)=c^2\sqrt{1+(3\pi^2\hbar^3\rho_e/m_e^4c^3)^{2/3}}, 
\end{equation}
we have, due to charge neutrality condition $\rho_e=\rho_pm_e/m_p$,
\begin{equation}
\left(\matrix{\delta\mu_n \cr \delta\tilde\mu \cr}\right)
=\left(\matrix{{\cal P}_{11} & {\cal P}_{12}\cr {\cal P}_{21} & {\cal P}_{22}\cr}\right)
\left(\matrix{\delta\rho_n \cr \delta\rho_p \cr}\right),
\label{def-chemi}
\end{equation}
where
\begin{equation}
{\cal P}_{11}=\left({\partial\mu_n\over \partial\rho_n}\right)_{\rho_p}, \quad
{\cal P}_{12}=\left({\partial\mu_n\over\partial\rho_p}\right)_{\rho_n}
=\left({\partial\tilde\mu\over\partial\rho_n}\right)_{\rho_p}={\cal P}_{21}, \quad
{\cal P}_{22}=\left({\partial\tilde\mu\over\partial\rho_p}\right)_{\rho_n}.
\end{equation}
We write the inverse of equation (\ref{def-chemi}) as
\begin{equation}
\left(\matrix{\delta\rho_n \cr \delta\rho_p \cr}\right)=
\left(\matrix{{\cal Q}_{11} & {\cal Q}_{12}\cr {\cal Q}_{21} & {\cal Q}_{22}\cr}\right)
\left(\matrix{\delta\mu_n \cr \delta\tilde\mu \cr}\right),
\end{equation}
where
\begin{equation}
\left(\matrix{{\cal Q}_{11} & {\cal Q}_{12}\cr {\cal Q}_{21} & {\cal Q}_{22}\cr}\right)
=\left(\matrix{{\cal P}_{11} & {\cal P}_{12}\cr {\cal P}_{21} & {\cal P}_{22}\cr}\right)^{-1}.
\end{equation}

The curl of the Euler equations, which is the vorticity equation, is useful 
for perturbation analysis of rotating stars, and is given in terms of $\delta V_n^a$ and 
$\delta V_p^a$ by 
\begin{eqnarray}
\epsilon^{abc}\nabla_b\left[
\left(i\omega\,\frac{\rho_{11}}{\rho_n}\,g_{cd}+2\Omega\,\nabla_d\varphi_c\right)\delta V_n^d  
+i\omega\frac{\rho_{12}}{\rho_n}\,g_{cd}\delta V_p^d \right]=0\,,
\label{curl-n}
\end{eqnarray}
\begin{eqnarray}
\epsilon^{abc}\nabla_b\left[
\left(i\omega\left(\frac{\rho_{22}}{\rho_p}+\frac{m_e}{m_p}\right)\,g_{cd}+
2\Omega\left(1+\frac{m_e}{m_p}\right)\nabla_d\varphi_c\right)\delta V_p^d 
+i\omega\frac{\rho_{12}}{\rho_p}\,g_{cd}\delta V_n^d \right]=0\,,
\label{curl-p}
\end{eqnarray}
where $g_{ab}$ and $\epsilon^{abc}$ mean the metric tensor and the Levi-Civita tensor 
in the flat space.

Although we have made no assumptions for rotation velocity up to now, let us restrict 
our consideration to slowly rotating stars in order to obtain inertial modes to the lowest 
order in the stellar rotation frequency $\Omega$. 
In the limit of slow rotation, then, we assume perturbations to obey 
the following ordering in $\Omega$: 
\begin{equation}
\delta V_n^a=O(1), \quad \delta V_p^a=O(1), \quad \omega/\Omega=O(1), \quad 
\delta\rho_n=O(\Omega), \quad \delta\rho_p=O(\Omega), \quad {\rm and} \quad 
\delta\Phi=O(\Omega)\,.
\label{scaling}
\end{equation}
Examining the zero-frequency subspace of the perturbation in a non-rotating superfluid 
core, Andersson \& Comer (2001) showed that solutions obeying the ordering  
(\ref{scaling}) may be allowed in the equations describing hydrodynamics in the 
superfluid core. The existence of solutions obeying (\ref{scaling}) is similar to 
that in a barotropic normal fluid star (Lockitch \& Friedman 1999). Those solutions 
can be interpreted as infinitely degenerate $g$-modes of zero-frequency if we consider 
no rotational effect. It is convenient to introduce dimensionless frequency $\kappa_0$ 
in the limit of $\Omega\rightarrow 0$, which is defined by
\begin{equation}
\kappa_0=\lim_{\Omega\rightarrow 0}\omega/\Omega\,.
\end{equation}
Then, basic equations describing the lowest order inertial modes are given by  
\begin{equation}
\nabla_a\,(\rho_n\delta V_n^a)=0, \quad \nabla_a\,(\rho_p\delta V_p^a)=0 ,
\label{con-eq}
\end{equation}
\begin{eqnarray}
q^a_n=\epsilon^{abc}\nabla_b\left[
\left(\frac{\rho_{11}}{\rho_n}\,g_{cd}-2i\kappa_0^{-1}\nabla_d\varphi_c\right)\delta V_n^d
+\frac{\rho_{12}}{\rho_n}\,g_{cd}\delta V_p^d \right]=0\,,
\label{curl-n2}
\end{eqnarray}
\begin{eqnarray}
q_p^a=\epsilon^{abc}\nabla_b\left[
\left(\left(\frac{\rho_{22}}{\rho_p}+\frac{m_e}{m_p}\right)\,g_{cd}-
2i\kappa_0^{-1}\,\left(1+\frac{m_e}{m_p}\right)\nabla_d\varphi_c\right)\delta V_p^d
+\frac{\rho_{12}}{\rho_p}\,g_{cd}\delta V_n^d \right]=0\,.
\label{curl-p2}
\end{eqnarray}
Note that $q^a_n$ ($q_p^a$) has only two independent components because $\nabla_a q^a_n=0$ 
($\nabla_a q^a_p$=0) is automatically satisfied.

>From similar consideration to that for the superfluid core, basic equations for pulsations  
in the normal fluid envelope are given by 
\begin{equation}
\nabla_a\,(\rho \delta v^a)=0\,,
\end{equation}
\begin{eqnarray}
q^a=\epsilon^{abc}\nabla_b\left[
\left(g_{cd}-2i\kappa_0^{-1}\nabla_d\varphi_c\right)\delta v^d \right]=0\,,
\end{eqnarray}
where $\delta v^a$ denotes the perturbation of the fluid velocity. 

The perturbations in a uniformly rotating star are expanded in terms of spherical harmonic
functions with different $l$'s for a given $m$ (e.g., Lee \& Saio 1986).
The vector $\delta V^a_X$ is therefore given by
\begin{equation}
\delta V_X^r=r \sum_{l\geq |m|} S_l^X(r)Y_l^m(\theta,\varphi)e^{i\sigma t},
\label{vr-ep}
\end{equation}
\begin{equation}
\delta V_X^\theta=\sum_{l\geq|m|}\left[H_l^X(r){\partial\over\partial\theta}
Y_l^m(\theta,\varphi)-i T_{l^\prime}^X(r){1\over\sin\theta}{\partial\over\partial\varphi}
Y_{l^\prime}^m(\theta,\varphi)\right]e^{i\sigma t},
\label{vt-ep}
\end{equation}
\begin{equation}
\delta V_X^\varphi={1\over\sin^2\theta}\sum_{l\geq|m|}\left[H_l^X(r){\partial\over
\partial\varphi}Y_l^m(\theta,\varphi)+i T_{l^\prime}^X(r)\sin\theta
{\partial\over\partial\theta}Y_{l^\prime}^m(\theta,\varphi)\right]e^{i\sigma t},
\label{vp-ep}
\end{equation}
where $l=|m|+2(k-1)$ and $l'=l+1$ for even modes, and $l=|m|+2k-1$ and $l'=l-1$
for odd modes, where $k=1,~2,~3,~\cdots$. Here, we have 
employed the constituent index $X$ which can be either $n$, $p$ or $o$, where $o$ means 
the velocity field in the normal fluid envelope, and the normal spherical polar 
coordinate $(r,\theta,\varphi)$ is used. In this paper, the even and odd modes are respectively 
characterized by symmetry and antisymmetry of their eigenfunctions with respect to the 
equatorial plan.  
Substituting these expansions into equations (\ref{con-eq}) through (\ref{curl-p2}), 
we obtain oscillation equations given as 
a set of simultaneous linear ordinary differential equations of the expansion 
coefficients (see Appendix), which is to be integrated in the superfluid core.
The oscillation equations solved in the normal fluid envelope are similar to those
given in Lockitch \& Friedman (1999).

To obtain a unique solution of an oscillation mode, solutions in
the superfluid core and in the normal fluid envelope are matched at the interface between 
the two domains by imposing jump conditions given by
\begin{equation}
\delta V^r_n=\delta V^r_p=\delta V^r_o \,.
\end{equation}
The boundary conditions at the stellar center is the regularity condition of the 
perturbations $\delta V^a_n$ and $\delta V^a_p$. The boundary condition at the 
stellar surface is $\Delta p=0$, where $\Delta p$ is the Lagrangian change in  
the pressure.

For numerical computation, oscillation equations of a finite dimension are obtained
by disregarding the terms
with $l$ larger than $l_{max}=|m|+2k_{max}-1$ in the expansions of perturbations
such as given by (\ref{vr-ep}) to (\ref{vp-ep}), where $l_{max}$ is determined so 
that the eigenfrequency and the eigenfunctions are well converged as $l_{max}$ 
increases. We solve the oscillation equations of a finite dimension as an eigenvalue 
problem with the scaled oscillation frequency $\kappa_0$ using a Henyey type relaxation 
method (see, e.g., Unno et al. 1989).

\section{Dissipations}

In the previous section, we consider no dissipation effects. Thus, the conserved 
energy $E$ of oscillations exits and is given by (Mendell 1991b) 
\begin{eqnarray}
E=&&{1\over 2}\int_{V_S} \left(\rho \delta v_a^*\delta v^a+
{\tilde\rho^2\over\rho}\delta{w}_a^*\delta{w}^a
+\sum_{i,j}{\cal P}_{ij}\delta\rho_i^*\delta\rho_j
-\frac{1}{4\pi G}\,\nabla_a\delta\Phi^*\nabla^a\delta\Phi
\right)\sqrt{g}\,d^3x \nonumber \\ 
&+&{1\over 2}\int_{V_N}\left(\rho\delta v_a^*\delta v^a
+\left(\frac{\delta p}{\rho}+\delta\Phi\right)\delta\rho^*\right)\sqrt{g}\,d^3x 
\end{eqnarray}
where $g$ is the determinant of the metric, and asterisk $(^*)$ denotes the complex 
conjugate of the quantity.  Here, $\int_{V_S}d^3x$ and $\int_{V_N}d^3x$ means the 
integration over the superfluid core and the normal fluid envelope, respectively. 
In this equation, we have employed new variables, defined by 
\begin{eqnarray}
&&\delta{v}^a=(\rho_n\delta v_n^a+\rho_p\delta v_p^a)/\rho=
(\rho_n\delta{V}_n^a+\rho_p\delta{V}_p^a)/\rho\,, \nonumber \\
&&\delta{w}^a=\delta v_n^a-\delta v_p^a=
{\rho_n\rho_p\over\tilde\rho^2}\left(\delta{V}_n^a-\delta{V}_p^a\right)\,, \\
&&\delta\rho_1=\delta\rho_n\,, \quad \delta\rho_2=\delta\rho_p\,. \nonumber
\label{def-v-w}
\end{eqnarray}
Note that $\rho\delta{v}^a$ is interpreted as the total momentum current vector of 
perturbation.  
In reality, however, there are a lot of dissipation mechanisms in real 
neutron stars. Thus, the oscillation energy $E$ is not conserved in real neutron stars. 
In other words, the amplitude of an oscillation mode should damp or grow depending 
on considered dissipation mechanisms.  
The stability of an oscillation mode of a star is then determined by 
summing up all contributions from various dissipation mechanisms.
In this study, we consider the contributions from gravitational radiation reaction, 
viscous processes, and mutual friction in the superfluid core. Then, the averaged 
energy loss (or gain) rate $dE/dt$ of a normal mode in a rotating neutron star may 
be given by 
\begin{eqnarray}
{dE\over dt}=&&-\sigma\omega\sum_{l'\ge |m|} N_l\sigma^{2l} 
\left|\delta J_{l'm}\right|^2 
-2\int_{V_S+V_N}\eta_s\delta\sigma^{ab*}\delta\sigma_{ab}\sqrt{g}\,d^3x 
\nonumber \\
&&-2\Omega\int_{V_S}\rho_nB_n\left({\tilde\rho^2\over\rho_n\rho_p}\right)^2
(g_{ab}-z_a z_b)\delta w^{a*}\delta w^b\sqrt{g}\,d^3x \nonumber\\
=&& \sum_{l'\ge |m|}\left({dE\over dt}\right)_{GJ,l'}+\left({dE\over dt}\right)_{S}
+\left({dE\over dt}\right)_{MF},
\label{dedt}
\end{eqnarray}
where $z^a$ is the unit vector that parallels the rotation axis, defined by 
$z_a=\epsilon_{abc}\nabla^b\varphi^c/2$.  

The terms $(dE/dt)_{GJ,l}$ on the right hand side of equation (\ref{dedt})
denote the energy loss (or gain) rates due to gravitational
radiation reaction associated with the mass current multipole moment $\delta J_{lm}$, 
where
\begin{equation}
\delta J_{lm}={2\over c(l+1)}
\int_{V_S+V_N} r^{l+1} \rho\delta{v}_a\epsilon^{abc}\nabla_bY_l^{m*}\nabla_cr\sqrt{g}\,d^3x
\end{equation}
and
\begin{equation}
N_l={4\pi G\over c^{2l+1}}{(l+1)(l+2)\over l(l-1)[(2l+1)!!]^2}, 
\end{equation}
and $c$ is the velocity of light (Thorne 1980, Lindblom et al. 1998).

The term $(dE/dt)_{S}$ is the energy dissipation rate due to shear viscosity, 
and $\eta_s$ is the shear viscosity coefficient, and $\delta\sigma_{ab}$ is 
the traceless rate of strain tensor for the perturbed velocity field, which is defined by 
\begin{equation}
\delta\sigma_{ab}=\frac{1}{2}\left(\nabla_a u_b+\nabla_b u_a
-\frac{2}{3}g_{ab}\nabla_c u^c\right)\,,
\end{equation}
for the velocity field $u^a$. 
In this paper, we ignore the contribution from the bulk viscosity, which is
important only for newly born hot neutron stars without superfluids in the core.
The shear viscosity coefficient we use in the superfluid core is
\begin{equation}
\eta_S=6.0\times 10^{18}\left({\rho\over 10^{15}{\rm g/cm^3}}\right)^2
\left({10^9 K\over T}\right)^2 {\rm g/cm ~s}
\end{equation}
(Cutler \& Lindblom 1987, Flowers \& Itoh 1979), and that in the normal fluid 
envelope is given by
\begin{equation}
\eta_S=1.95\times 10^{18}\left({\rho\over 10^{15}{\rm g/cm^3}}\right)^{9/4}
\left({10^9 K\over T}\right)^2 {\rm g/cm ~s}
\end{equation}
(Cutler \& Lindblom 1987, Flowers \& Itoh 1979).
The rate of strain $\delta\sigma_{ab}$ is evaluated by using $\delta{v}^a_e$ 
in the superfluid core, and $\delta{v}^a$ in the normal fluid envelope 
(Lindblom \& Mendell 2000).

The term $(dE/dt)_{MF}$ is the energy loss rate due to 
the mutual friction in the superfluid core, 
and the dimensionless coefficient $B_n$ is given by (Mendell 1991b)
\begin{equation}
B_n=0.011\times {\rho_p\over\rho_n}
\left({\rho_{pp}\over\rho_{p}}\right)^{1/2}
\left({\rho_{pn}\over\rho_{pp}}\right)^2
\left({\rho_p\over 10^{14}{\rm g/cm^3}}\right)^{1/6}.
\end{equation}
The mutual friction is a dissipation mechanism inherent to a rotating system 
of superfluids, and it is caused by scattering of normal fluid particles off the vortices
in the superfluids.
Since we have assumed perfect charge neutrality between the electrons and protons, we consider
scattering between the normal electrons and vortices of the neutron superfluid
(Mendell 1991b, see also Alpar, Langer, \& Sauls, 1984).

As is indicated by the first term on the right hand side of equation (\ref{dedt}), 
if the normal mode has an oscillation frequency that satisfies
$-\sigma\omega>0$, 
the oscillation energy $E$ in the corotating frame, in the absence of other damping mechanisms, 
increases as a result of
gravitational wave radiation, indicating instability of the mode (Friedman \& Schutz 1978).

The damping (or growth) time-scale $\tau$ of a normal mode may be given by
\begin{equation}
{1\over\tau}=-{1\over 2E}\left({dE\over dt}\right)=\sum_{l'>|m|}{1\over\tau_{GJ,l'}}+
{1\over\tau_S}+{1\over\tau_{MF}},
\end{equation}
where $\tau_i=-2E/(dE/dt)_i$. For the inertial modes, it is convenient to
derive an extrapolation formula of the time-scale $\tau$ given as a function of $\Omega$
and the interior temperature $T$ (e.g., Lindblom et al. 1998, Lindblom \& Mendell 2000):
\begin{equation}
{1\over\tau (\Omega,T)}
=\sum_{l'\ge |m|}{1\over\tau^0_{GJ,l'}}\left({\Omega^2\over \pi G\bar\rho}\right)^{l'}
+{1\over\tau^0_S}\left({10^9 K\over T}\right)^2
+{1\over\tau^0_{MF}}\left({\Omega^2\over \pi G\bar\rho}\right)^{1/2},
\end{equation}
where $\bar\rho=M/(4\pi R^3/3)$. Here, $M$ and $R$ are the mass and radius of the star. 
The quantities $\tau^0_{GJ,l'}$, $\tau^0_S$, and $\tau^0_{MF}$ are assumed to be 
independent on $\Omega$ and $T$.

\section{Numerical Results}

Following Lindblom \& Mendell (2000), we employ a polytropic model of 
index $N=1$ with the mass $M=1.4M_\odot$ and the radius $R=12.57$km
as a background model for modal analysis. The model is divided into 
a superfluid core and a normal fluid envelope, the interface of which 
is set at $\rho=\rho_s=2.8\times 10^{14}$g/cm$^3$. In the normal fluid 
envelope, we use the polytropic equation of state given by $p=K\rho^2$
both for the equilibrium structure and for the oscillation equations.
The core is assumed to be filled with neutron and proton superfluids 
and a normal fluid of electron, for which an equation of state, labeled
A18+$\delta v$+UIX (Akmal, Pandharipande, \& Ravenhall 1998), is used to 
give the thermodynamical quantities used for the oscillation equations.
For the mass density coefficients $\rho_{nn}$, $\rho_{pp}$, and 
$\rho_{np}$ in the core, we employ an empirical relation given by 
(see Lindblom \& Mendell 2000)
\begin{equation}
{\rho_p/\rho}\approx 0.031+8.8\times 10^{-17}\rho, 
\end{equation}
and a formula given by
\begin{equation}
\rho_{np}=-\eta\rho_n, 
\end{equation}
where $\eta$ is a parameter 
of order of $\sim 0.04$ (Borumand, Joynt, \& Klu\'zniak 1996).
In this paper we call $\eta$ the entrainment parameter.
In this study, only the modes having $m=2$ are examined because they are the 
most significant for the gravitational radiation driven instability. 
For the range of the entrainment parameter $\eta$, we choose 
$-0.01\le\eta\le 0.04$. 
Although negative values of $\eta$ might not be 
suitable for physical models of neutron stars, we include in our calculations negative $\eta$ to 
demonstrate clearly the $\eta$ dependence of the inertial modes around $\eta\sim 0$.

We find that inertial modes in the superfluid core are split 
into two families, which we call ordinary fluid inertial modes ($i^o$-mode), and 
superfluid inertial modes ($i^s$-mode). 
This mode splitting is 
common in other oscillation modes in the superfluid core, 
because the dynamical degree of freedom in the 
superfluid core simply doubles 
(see, e.g., Epstein 1988; Lindblom \& Mendell 1994; Lee 1995; Andersson \& Comer 2001; Prix \& 
Rieutord 2002). 
As a subclass of inertial modes, the $r$-modes of neutron stars with a superfluid core are
also split into ordinary fluid-like $r$-modes and superfluid-like $r$-modes,
which we respectively call $r^o$-modes and $r^s$-modes as in Lee \& Yoshida (2002).
Since basic properties of the eigenfunctions
$\delta v^a$ (or $\delta w^a$) of an inertial mode
are quite similar to the eigenfunctions $\delta v^a$ of the corresponding inertial mode in 
the normal fluid star without core superfluidity,
we employ the same classification scheme as that employed by
Lockitch \& Friedman (1999) and Yoshida \& Lee (2000a) for inertial modes in the normal fluid star.
In this scheme,
inertial modes are classified in terms of the two quantum numbers $m$ and $l_0-|m|$, and 
the scaled frequency $\kappa_0\equiv\lim_{\Omega\rightarrow 0}\omega/\Omega$.
For given $m$ and $l_0-|m|$, the number of inertial modes of one kind is $l_0-|m|$.
Even parity inertial modes have even $l_0-|m|$,  
and odd parity inertial modes have odd $l_0-|m|$.
Note that the $r$-modes are classified as inertial modes with $l_0-|m|=1$ (e.g., Yoshida \& Lee 2000a).
It is interesting to note that,
although there exist three quantum numbers to classify the normal modes of non-rotating stars,
a third quantum number may be missing for inertial modes.
Usually, the ratio $E_S/E$ of the oscillation energy $E_S$ in the core 
to the total oscillation energy $E$
is a good indicator for a distinction between $i^s$- and $i^o$-modes, since we find
$E_S/E\ltsim 0.5$ for the $i^o$-modes and $E_S/E\gtsim0.9$ for the $i^o$-modes for the models 
composed of a superfluid core and a normal fluid envelope.

Let us first consider the case of $\eta=0$, that is, the case of no entrainment.
In Table 1, $\kappa_0$'s for $i^o$- and $i^s$-modes of low $l_0-|m|$
are tabulated together with $\kappa_0$'s for the corresponding inertial modes in the 
$N=1$ polytrope model without superfluidity in the core, which we simply call $i$-mode 
(see, Lockitch \& Friedman 1999; Yoshida \& Lee 2000a). 
Our values of $\kappa_0$ for the $i^o$- and $i^s$-modes are in good 
agreement with those computed by Lee \& Yoshida (2002) with a different method. 
Note that we have shown five digits for numerical values of $\kappa_0$ 
to indicate that our numerical 
results include at most a few percent numerical error.  
We can see that $\kappa_0$'s of paired $i^o$- and $i^s$-modes have quite similar 
values to that of the corresponding inertial mode in the $N=1$ polytropic normal fluid model. 
In order to demonstrate model properties of the inertial modes, we 
exhibit the expansion coefficients $S^v_l$, $T^v_{l'}$, $S^w_l$, and $T^w_{l'}$
of the eigenfunctions $\delta v^a$ and $\delta w^a$, defined by 
\begin{equation}
S_l^v=(\rho_nS_l^n+\rho_pS_l^p)/\rho\,,\quad 
T_{l'}^v=(\rho_nT_{l'}^n+\rho_pT_{l'}^p)/\rho\,,
\end{equation}
\begin{equation}
S_l^w=\rho_n\rho_p(S_l^n-S_l^p)/\tilde\rho^2\,,\quad 
T_{l'}^w=\rho_n\rho_p(T_{l'}^n-T_{l'}^p)/\tilde\rho^2\,.
\end{equation}
Note that $S_l^w$ and $T_{l'}^w$ vanish in the normal fluid envelope because 
neutrons and protons co-move. 
In Figures 1 through 4, the first three expansion coefficients $S^v_l$, $T^v_{l'}$, 
$S^w_l$, and $T^w_{l'}$ are shown for the $i^o$- and $i^s$-modes of $l_0-|m|=2$ which have
$\kappa_0=-0.5685$ and $\kappa_0=-0.5115$, respectively. 
Here, the amplitude normalization given by $T^v_{m+1}=1$ has been used.
It is noted that 
$\delta v^a$ of the $i^o$-modes and $\delta w^a$ of the $i^s$-modes 
resemble $\delta v^a$ of the corresponding inertial modes of the normal fluid star model
(compare the figures with Figure 1 of Yoshida \& Lee 2000a). 
It is also found that although the amplitudes of $\delta v^a$ and $\delta w^a$
are similar to each other for the $i^o$-modes, the amplitude of 
$\delta w^a$ dominates that of $\delta v^a$ for the $i^s$-modes, which indicates that
the two superfluids in the core basically counter-move for the $i^s$-modes. 

In Figures 5 to 8, $\kappa_0$'s of inertial modes of low $l_0-|m|$ are plotted
as functions of the entrainment parameter $\eta$, where 
Figure 5 is for even parity modes with positive $\kappa_0$, 
Figure 6 for even parity modes with negative $\kappa_0$,
Figure 7 for odd parity modes with positive $\kappa_0$, and
Figure 8 for odd parity modes with negative $\kappa_0$.
In Figure 7, we have also given $\kappa_0$ of the $r^s$-mode, 
the locus of which as a function of $\eta$ is
in good agreement with the locus calculated by Lee \& Yoshida (2002) with a different method. 
Note that we cannot obtain $r^o$-modes by using our numerical method, the reason for which
is not clear (see the next section)
\footnote{Recently, we have generalized our formalism to the case of general relativistic
superfluid neutron stars.
In that study, our numerical code for Newtonian dynamics have been straightforwardly 
extended into the one for general relativistic dynamics. 
With this relativistic version of the 
code, both $r^o$- and $r^s$-modes can be obtained even in almost the Newtonian 
regime with small values of the relativistic parameter $M/R$. 
Fundamental differences in the two formalisms 
reside only in the master equations. 
As shown in equations (49)--(52), the basic equations 
for neutron and proton fluids in the Newtonian case get to degenerate 
in lowest order of $\Omega$ when $r^o$-modes are considered, while the basic equations
in the relativistic case don't
even if we are concerned with $r^o$-modes (see, Yoshida \& Lee 2003). 
We therefore guess that the reason why $r^o$-modes 
cannot be obtained in the Newtonian case is this possible degeneracy of 
the two sets of superfluid equations in Newtonian dynamics.}.
From Figures 5 through 8, we find that, although $\kappa_0$ for the $i^o$-modes 
is only weakly dependent on the parameter $\eta$, 
$|\kappa_0|$ for the $i^s$-modes increases almost linearly as 
$\eta$ increases. 
Because of these different dependences of $\kappa_0$ on the parameter $\eta$, 
frequent avoided crossings are inevitable between $i^o$-modes and $i^s$-modes. 
As shown by Figure 7, for example, the $i^s$-mode of $l_0-|m|=3$ 
which has $\kappa_0=1.3865$ at $\eta=0$ 
suffers three avoided crossings with different $i^o$-modes 
as $\eta$ is increased from $-0.01$ to $0.04$.  
Since $\kappa_0$'s of paired $i^o$- and $i^s$-modes 
at $\eta=0$ are close to each other, they always make an 
avoided crossing around $\eta=0$, which is also shown by Figures 5 through 8.

In Figures 9 through 12, we display the first three coefficients $S^v_l$, 
$T^v_{l'}$, $S^w_l$, and $T^w_{l'}$ for the $i^o$- and $i^s$-modes of $l_0-|m|=2$, 
which respectively have $\kappa_0=-0.5403$ and $\kappa_0=-0.6103$ at $\eta=0.01$. 
Here, the amplitude normalization given by $T^v_{m+1}=1$ has been used. 
Note that these two inertial modes have $\kappa_0=-0.5685$ and $\kappa_0=-0.5115$ at $\eta=0$.
Comparing these figures with Figures 1 to 4, we observe that 
the basic properties of the eigenfunctions 
of the inertial modes do not depend on the entrainment parameter $\eta$, except at avoided crossings.

In Table 2, we tabulate $\kappa_0$ and the
scaled damping (or growth) timescales $\tau^0_i$ for the $i^o$- and $i^s$-modes with positive $\kappa_0$
for $\eta=0$, $0.02$, and $0.04$.  
We find that the growth timescales $\tau^0_{GJ,l'}$ for all the 
$i^o$- and $i^s$-modes in the table are longer than that of the $r$-mode of the normal fluid neutron star, 
which means that the instability of the $i^o$- and $i^s$-modes caused by the current multipole radiation 
is not strong. 
This result is the same as that for the inertial modes in a normal fluid star. 
It is interesting to note that the two $i^s$-modes in the table having $\kappa_0=2.1661$ and 
$\kappa_0=2.4083$ at $\eta=0.04$ are stable
against gravitational radiation reaction since they have positive $\tau^0_{GJ,l'}$.  
The reason for this can be understood as follows.
As mentioned in the last section, the gravitational radiation driven instability sets in when the 
frequency of the mode satisfies $\sigma(\sigma-m\Omega) <0$, which leads to
the instability criterion given by $0<\kappa_0<m$ for positive $m$.
For inertial modes of normal fluid stars, it is known that
all inertial modes having positive $\kappa_0$ for $m>0$ satisfy the instability criterion
(Lindblom \& Ipser 1999; Lockitch \& Friedman 1999; Yoshida \& Lee 2000a). 
For inertial modes in the superfluid core, on the other hand, 
since the frequency $|\kappa_0|$ of the 
$i^s$-modes monotonically 
increases as the value of $\eta$ is increased, the $\kappa_0$ of some positive-frequency 
$i^s$-modes exceeds the limit $m$ when $\eta$ gets larger than a critical value $\eta_{crit}$, 
the value of which depends on the mode (see, Figures 5 and 7). Thus, some of 
the positive-frequency $i^s$-modes that are unstable satisfying the instability criterion
at $\eta=0$ become stable invalidating the criterion for $\eta$ larger than $\eta_{crit}$. 
As for the effects of mutual friction, we observe that the damping timescales 
$\tau^0_{MF}$ for the inertial modes are much shorter than those due to other 
dissipation processes unless mutual
friction between the superfluids is extremely weak, that is, the parameter $\eta$
is extremely small.
Note that the amount of dissipation caused by the mutual 
friction is roughly proportional to $B_n|\delta w^a|^2$ where $B_n\propto\eta^2$.
Accordingly, if we assume $\eta\sim 0.04$ as a typical value for the parameter,
the dissipation due to mutual friction in the superfluid core is very effective to 
damp out the $i^o$- and $i^s$-modes.

\section{Discussion}

As mentioned in the previous section, with our method of calculation
we fail in obtaining ordinary fluid $r$-modes, which are called $r^o$-modes
in Lee \& Yoshida (2002). 
However, this does not necessarily mean that our basic equations (A2)--(A7) do not contain
such $r$-mode solutions. 
To prove this, let us consider purely axial parity 
solutions, whose angular behavior is given by a purely axial harmonic function having $l'=m$. 
In this section, we also assume for simplicity that the densities $\rho_n$, $\rho_p$, and $\rho_{12}$
are constant and do not depend on $r$.
In virtue of the assumption of purely axial 
mode, we can reduce our basic equations to the following equations (see, Appendix): 

\noindent
For the superfluid core, 
\noindent
\begin{equation}
\nu m(m+2)J^m_{m+1}\left\{r{d\over dr} T^n_m-(m-1)T^n_m\right\}=0\,, 
\label{r-eq1}
\end{equation}
\begin{equation}
\nu m(m+2)J^m_{m+1}\left\{r{d\over dr} T^p_m-(m-1)T^p_m\right\}=0\,,
\label{r-eq2}
\end{equation}
\begin{equation}
\left(1-{m\nu\over m(m+1)}\right)T^n_m+{\rho_{12}\over \rho_n}(T^p_m-T^n_m)= 0\,,
\label{r-eq3}
\end{equation}
\begin{equation}
\left(1-{m\nu\over m(m+1)}\right)T^p_m+
{\rho_{12}\over \rho_p}(T^n_m-T^p_m)= 0\,,
\label{r-eq4}
\end{equation}

\noindent
For the normal fluid envelope, 
\noindent
\begin{equation}
\nu m(m+2)J^m_{m+1}\left\{r{d\over dr} T^o_m-(m-1)T^o_m\right\}=0\,,         
\label{r-eq5}
\end{equation}
\begin{equation}
\left(1-{m\nu\over m(m+1)}\right)T^o_m= 0\,,
\label{r-eq6}
\end{equation}
where $\nu=2\Omega/\sigma$ and $J^m_l=[(l^2-m^2)/(4l^2-1)]^{1/2}$. 
Here, we have ignored the terms due to $m_e/m_p$ for simplicity. 
The condition for the continuity of the total 
momentum flux at the interface between the superfluid core and the 
normal fluid envelope is given by
\begin{equation}
T^o_m=(\rho_nT^n_m+\rho_pT^p_m)/\rho \quad {\rm at}\ r=r_s\,,
\end{equation}
where $r_s$ is the radius of the superfluid core.
Because the factor $\nu m(m+2)J^m_{m+1}$ in equations 
(\ref{r-eq1}), (\ref{r-eq2}) and (\ref{r-eq5}) does not vanish for $\nu\not=0$ and $m\ge 2$, 
we can obtain solutions to differential equations (\ref{r-eq1}), (\ref{r-eq2}), 
and (\ref{r-eq5}) as 
\begin{equation}
T^n_m=A_n (r/r_s)^{m-1}\,, \quad  T^p_m=A_p (r/r_s)^{m-1}\,, \quad  
T^o_m=B (r/r_s)^{m-1}\,,
\label{sol1}
\end{equation}
where $A_n$, $A_p$, 
and $B$ are constants, which are to be determined by boundary condition (55) and constraint 
equations (51), (52), (54). 
Substituting the solutions (56) into algebraic equations (51), (52), (54) and (55), we obtain
\begin{equation}
\left(\matrix{f-\rho_{12}/\rho_n & \rho_{12}/\rho_n \cr
\rho_{12}/\rho_p & f-\rho_{12}/\rho_p \cr}\right)\left(\matrix{A_n \cr A_p \cr}\right)
\equiv W\left(\matrix{A_n \cr A_p \cr}\right)=0,
\end{equation}
\begin{equation}
fB=0,
\end{equation}
and
\begin{equation}
B=(\rho_n(r_s)A_n+\rho_p(r_s)A_p)/\rho(r_s)\,,
\label{con1}
\end{equation}
where $f=1-\nu/(m+1)$. 
For non-trivial solutions to exist in the core, we must have $\det W=0$, which leads to
$f=0$ or $f=\rho_{12}(\rho_n+\rho_p)/(\rho_n\rho_p)$.
For the case of $f=0$ (i.e. $\kappa_0=2/(m+1)$), solutions may be given by
\begin{equation}
A_n=A_p=B\,.
\label{sol1a}
\end{equation}
This solution obviously corresponds to the $r^o$-mode since
the oscillation frequency $\kappa_0$ 
is exactly the same as that of the $r^o$-mode, and the two superfluids flow together in the core. 
For the case of $f=\rho_{12}(\rho_n+\rho_p)/(\rho_n\rho_p)$ (i.e., $\kappa_0\not=2/(m+1)$), 
on the other hand, solutions are given by
\begin{equation}
(\rho_n(r_s)A_n+\rho_p(r_s)A_p)/\rho(r_s)=B=0\,.
\label{con1}
\end{equation}
This solution corresponds to the $r^s$-modes found by Lee \& Yoshida (2002), for which
the two superfluids counter-move in the core (see, also, Andersson \& Comer 2001).
It is to be noted that the solution (60) for arbitrary $\eta$ and the solution (61) for
$\eta=0$ can be derived even without the assumption that the densities $\rho_n$, $\rho_p$,
and $\rho_{12}$ do not depend on the radial distance $r$.

\section{Conclusion}

We have discussed the modal properties of inertial modes of rotating neutron stars
with the core filled with neutron and proton superfluids, taking account of entrainment
effects between the superfluids. We find that the inertial modes 
are split into two families, ordinary fluid inertial modes ($i^o$-mode) and 
superfluid inertial modes ($i^s$-mode). 
The two superfluids in the core counter-move for the $i^s$-modes. 
For the $i^o$-modes, the dimensionless frequency $\kappa_0$ is only weakly
dependent on the entrainment parameter $\eta$. For the $i^s$-modes, on the other hand, 
$|\kappa_0|$ almost linearly increases as the entrainment parameter $\eta$ is increased.
Because of the different dependences of $\kappa_0$ on the parameter $\eta$,
avoided crossings as functions of $\eta$ are common between $i^o$- and $i^s$-modes.
We also find that some of the $i^s$-modes 
that are unstable against the gravitational radiation reaction at $\eta=0$ 
become stable invalidating the instability criterion when $\eta$ becomes larger than some critical 
$\eta_{crit}$, the value of which depends on the mode.
The radiation driven instability associated with the current 
multipole radiation is quite weak for the inertial modes. 
Since the mutual friction damping in the superfluid core is quite strong for a typical value of $\eta$ 
compared with other damping mechanisms considered in this paper, we conclude that the inertial modes are
easily damped by the mutual friction.

Recently, Arras et al. (2002) considered non-linear couplings between an unstable $r$-mode
and stable inertial modes of rotating normal fluid neutron stars as a mechanism
that limits the amplitude growth of the $r$-mode, and discussed in the WKB limit 
the modal properties of the inertial modes, which
are important to determine the saturation amplitude of the $r$-mode.
In their analysis, however, no effects of buoyancy and possible core superfluidity on the 
inertial modes are correctly included, which substantially affect the modal properties such
as the frequency spectra and damping rates (see Yoshida \& Lee 2000b).
If we consider cold neutron stars in a neutron star binary, inertial modes in 
the superfluid core of the stars will be excited by the tidal interaction with the companion, 
and the mutual friction dissipation produced by the excited inertial modes will play an important 
role in the energy and angular momentum transfer between the orbital motion and 
the neutron stars, since no $g$-modes are propagating in the superfluid core (Lee 1995, 
Andersson \& Comer 2001). 
In fact, 
the damping timescale of the inertial modes propagating in the superfluid core may be
given by $\tau\sim\tau^0_{MF}/\bar\Omega$ sec, and it is for $\bar\Omega\sim0.1$ 
of order of 0.1 to 1 sec, which is extremely short compared with the damping timescales
of $g$-modes due to viscous processes and emission of gravitational waves
(see, e.g., McDermott, Van Horn, \& Hansen 1988; Reisenegger \& Goldreich 1992).
Considering that the orbital period decrease $d\ln P/dt$ due to gravitational radiation 
reactions is proportional to $D^{-4}$ (e.g., Shapiro \& Teukolsky 1983) but 
that due to the tidal interaction to $D^{-5}$ (e.g., Savonije \& Papaloizou 1983)
where $D$ is the binary separation,
the tidal interaction in terms of the inertial modes in the superfluid core becomes
important as the binary separation decreases.

\vskip 0.5cm
\noindent{\sl  Acknowledgments:} 
{S.Y. acknowledges financial support from Funda\c c\~ao para a  Ci\^encia e a Tecnologia 
(FCT) through project SAPIENS 36280/99.}
\vskip 0.5cm

\appendix

\section{Oscillation Equations}

For the oscillation equations in the superfluid core,
we employ vectors $\pmb{s}^n$, $\pmb{s}^p$, $\pmb{t}^n$, $\pmb{t}^p$,
$\pmb{h}^n$, and $\pmb{h}^p$, whose components are given by
\begin{equation}
s_{k}^n=S_l^n, \quad s_{k}^p=S_l^p, \quad
t_{k}^n=T_{l'}^n, \quad t_{k}^p=T_{l'}^p, \quad
h_{k}^n=H_l^n, \quad h_{k}^p=H_l^p,
\end{equation}
where $l=|m|+2(k-1)$, $l^\prime=l+1$ for even modes and $l=|m|+2k-1$, 
$l^\prime=l-1$ for odd modes, where $k=1,~2,~3, \cdots$.
Using these vectors,
the perturbed continuity equations (\ref{con-n}) and (\ref{con-p}) are written as
\begin{eqnarray}
r{d\pmb{s}^n\over dr}=-\left({d\ln\rho_n\over d\ln r}+3\right)\pmb{s}^n
+\Lambda_0\pmb{h}^n\,,
\end{eqnarray}
\begin{eqnarray}
r{d\pmb{s}^p\over dr}=-\left({d\ln\rho_n\over d\ln r}+3\right)\pmb{s}^p
+\Lambda_0\pmb{h}^p\,.
\end{eqnarray}
The independent components of equations (\ref{curl-n2}) and (\ref{curl-p2}) are reduced to
\begin{eqnarray}
&&L_0r{d\pmb{h}^n\over dr}-M_1r{d\pmb{t}^n\over dr}
-m\nu\Lambda_0^{-1}r{d\pmb{s}^n\over dr}+
\{(2+m\nu)\pmb{1}-2m\nu\Lambda_0^{-1}\}\pmb{h}^n-
(\pmb{1}+2m\nu\Lambda_0^{-1})\pmb{s}^p \nonumber \\ 
&&+\nu\Lambda_0^{-1}C(\Lambda_1-2\pmb{1})\pmb{t}^n
+\frac{\rho_{12}}{\rho_n}\,\left\{ r{d\over dr}(\pmb{h}^p-\pmb{h}^n)+
2(\pmb{h}^p-\pmb{h}^n)-(\pmb{s}^p-\pmb{s}^n)\right\}  \\
&&+r{d\over dr}\left(\frac{\rho_{12}}{\rho_n}\right)\,(\pmb{h}^p-\pmb{h}^n)=0\,,
\nonumber
\end{eqnarray}
\begin{eqnarray}
&&L_0r{d\pmb{h}^p\over dr}-M_1r{d\pmb{t}^p\over dr}
-m\nu\Lambda_0^{-1}r{d\pmb{s}^p\over dr}+
\{(2+m\nu)\pmb{1}-2m\nu\Lambda_0^{-1}\}\pmb{h}^p-
(\pmb{1}+2m\nu\Lambda_0^{-1})\pmb{s}^p \nonumber \\
&&+\nu\Lambda_0^{-1}C(\Lambda_1-2\pmb{1})\pmb{t}^p
+\frac{\rho_{12}}{\rho_p}\left(1+\frac{m_e}{m_p}\right)^{-1}\,
\left\{ r{d\over dr}(\pmb{h}^n-\pmb{h}^p)+
2(\pmb{h}^n-\pmb{h}^p)-(\pmb{s}^n-\pmb{s}^p)\right\} \\
&&+r{d\over dr}\left\{\frac{\rho_{12}}{\rho_p}\left(1+\frac{m_e}{m_p}\right)^{-1}\right\}
\,(\pmb{h}^n-\pmb{h}^p)=0\,, \nonumber
\end{eqnarray}
\begin{equation}
L_1\pmb{t}^n-M_0\pmb{h}^n+\nu K\pmb{s}^n
+\frac{\rho_{12}}{\rho_n}\,(\pmb{t}^p-\pmb{t}^n)=0\,,
\end{equation}
\begin{eqnarray}
L_1\pmb{t}^p-M_0\pmb{h}^p+\nu K\pmb{s}^p
+\frac{\rho_{12}}{\rho_p}\,\left(1+\frac{m_e}{m_p}\right)^{-1}(\pmb{t}^n-\pmb{t}^p)=0\,, 
\end{eqnarray}
where $\nu=2\Omega/\sigma=2/\kappa_0$ and $\pmb{1}$ stands for a unit matrix. Here, 
the quantities, $L_0$, $L_1$, $C$, $K$, $\Lambda_0$, $\Lambda_1$, $M_0$, and $M_1$
are matrices, given as follows: 

\noindent
For even modes,
\[
({L_0})_{i,i} = 1-{m\nu\over l(l+1)} \, , \hspace{.3in}
({L_1})_{i,i} = 1-{m\nu\over (l+1)(l+2)} \, ,
\]
\[
({C})_{i,i} = - (l+2) J^m_{l+1} \, , \hspace{.3in}
({C})_{i+1,i} = (l+1) J^m_{l+2} \, ,
\]
\[
({K})_{i,i} = \frac{J^m_{l+1}}{l+1} \, , \hspace{.3in}
({K})_{i,i+1} = - \frac{J^m_{l+2}}{l+2} \, ,
\]
\[
({\Lambda}_0)_{i,i} = l(l+1) \, , \hspace{.3in}
({\Lambda}_1)_{i,i} = (l+1)(l+2) \, ,
\]
\[
({M}_0)_{i,i} = \nu\frac{l}{l+1} \, J^m_{l+1} \, , \hspace{.3in}
({M}_0)_{i,i+1} = \nu\frac{l+3}{l+2} \, J^m_{l+2} \, ,
\]
\[
({M}_1)_{i,i} = \nu\frac{l+2}{l+1} \, J^m_{l+1} \, , \hspace{.3in}
({M}_1)_{i+1,i} = \nu\frac{l+1}{l+2} \, J^m_{l+2} \, ,
\]
where $l=\vert m \vert + 2 i -2 $ for $i = 1,2,3,\dots $, and
\begin{equation}
J^m_l \equiv \left[ \frac{(l+m)(l-m)}{(2l-1)(2l+1)} \right]^{1/2}
\, .  \label{J_l}
\end{equation}
For odd modes,
\[
({L_0})_{i,i} = 1-{m\nu\over l(l+1)} \, , \hspace{.3in}
({L_1})_{i,i} = 1-{m\nu\over l(l-1)} \, ,
\]
\[
({C})_{i,i} = (l-1) J^m_{l} \, , \hspace{.3in}
({C})_{i,i+1} = -(l+2) J^m_{l+1} \, ,
\]
\[
({K})_{i,i} = - \frac{J^m_l}{l} \, , \hspace{.3in}
({K})_{i+1,i} = \frac{J^m_{l+1}}{l+1} \, ,
\]
\[
({\Lambda}_0)_{i,i} = l(l+1) \, , \hspace{.3in}
({\Lambda}_1)_{i,i} = l(l-1) \, ,
\]
\[
({M}_0)_{i,i} = \nu\frac{l+1}{l} \, J^m_l \, , \hspace{.3in}
({M}_0)_{i+1,i} = \nu\frac{l}{l+1} \, J^m_{l+1} \, ,
\]
\[
({M}_1)_{i,i} = \nu\frac{l-1}{l} \, J^m_l \, , \hspace{.3in}
({M}_1)_{i,i+1} = \nu\frac{l+2}{l+1} \, J^m_{l+1} \, ,
\]
where $l=\vert m \vert + 2 i -1 $ for $i = 1,2,3,\dots $.

The oscillation equations in the superfluid core which we use are given as a set of 
linear ordinary differential equations for the variables $\pmb{s}^X$ and $\pmb{t}^X$ 
with $X=n, p$, which are obtained by eliminating the vectors $\pmb{h}^X$ in 
equations (A2)--(A5) using the algebraic equations (A6) and (A7).
The oscillation equations in the normal fluid envelope can be obtained by substituting 
$\pmb{s}^n=\pmb{s}^p=\pmb{s}^o$, $\pmb{t}^n=\pmb{t}^p=\pmb{t}^o$ and 
$\rho_n=\rho_p=\rho$ into the 
oscillation equations in the superfluid core, where $\pmb{s}^o$ and $\pmb{t}^o$ means 
the fluid velocity fields in the normal fluid envelope.

\newpage 

\setcounter{section}{0}

\begin{table*}
\caption{$\kappa_0$ for the $i^o$- and $i^s$-modes for the $\eta=0$ model}
\begin{tabular}{crrr}
\hline
\hline
$l_0-|m|$ & $i^o$-mode & $i^s$-mode & $i$-mode \\
\hline
1 &-$\quad$&-$\quad$& 0.6667 \\
2 & 1.0978 & 1.1187 & 1.1000 \\
  &-0.5685 &-0.5115 &-0.5566 \\
3 & 1.3564 & 1.3865 & 1.3578 \\
  & 0.5180 & 0.5078 & 0.5173 \\
  &-1.0293 &-0.9729 &-1.0259 \\
4 & 1.5191 & 1.5522 & 1.5196 \\
  & 0.8639 & 0.8611 & 0.8630 \\
  &-0.2738 &-0.2421 &-0.2753 \\
  &-1.2738 &-1.2242 &-1.2729 \\
\hline
\end{tabular}
\end{table*}

\begin{table*}
\caption{\label{coefficients0}$\kappa_0$ and $\tau^0_i$'s for the 
$i^o$- and $i^s$-modes of $m=2$}
\begin{tabular}{cccccccccc}
\hline
\hline
Modes & $l_0-|m|$ & $\eta$ & $\kappa_0$ & $\tau^0_{GJ,2}$(s) & $\tau^0_{GJ,3}$(s) &
$\tau^0_{GJ,4}$(s) & $\tau^0_{GJ,5}$(s) & $\tau^0_{S}$(s) & $\tau^0_{MF}$(s) \\
\hline
$i^o$& 2 & 0    & 1.0978 &            &$-4.1\times 10^4$ &                        &
$-2.4\times 10^{14}$ &$2.5\times 10^6$&$\infty$ \\
$i^s$&   &      & 1.1187 &            &$-2.5\times 10^5$ &                        &
$-1.2\times 10^{15}$ &$1.2\times 10^6$&$\infty$ \\
$i^o$& 3 &      & 1.3564 &$-2.2\times 10^8$ &            &$-1.7\times 10^9$ &
                     &$3.3\times 10^6$&$\infty$ \\
$i^s$&   &      & 1.3865 &$-3.7\times 10^9$ &            &$-3.5\times 10^{10}$ &
                     &$5.2\times 10^5$&$\infty$ \\
$i^o$&   &      & 0.5180 &$-4.5\times 10^6$ &            &$-2.1\times 10^6$ &
                     &$2.6\times 10^6$&$\infty$ \\
$i^s$&   &      & 0.5078 &$-5.9\times 10^7$ &            &$-1.8\times 10^7$ &
                     &$5.7\times 10^5$&$\infty$ \\
$i^o$& 4 &      & 1.5191 &            &$-2.1\times 10^{11}$&                      &
$-2.4\times 10^{14}$ &$3.8\times 10^6$&$\infty$ \\
$i^s$&   &      & 1.5522 &            &$-5.9\times 10^{12}$&                      &
$-1.5\times 10^{16}$ &$2.2\times 10^5$&$\infty$ \\
$i^o$&   &      & 0.8639 &            &$-1.8\times 10^7$   &                      &
$-1.5\times 10^{10}$ &$9.9\times 10^6$&$\infty$ \\
$i^s$&   &      & 0.8611 &            &$-2.0\times 10^8$   &                      &
$-5.0\times 10^{11}$ &$2.3\times 10^6$&$\infty$ \\
$i^o$& 2 & 0.02 & 1.0912 &            &$-1.5\times 10^7$ &                        &
$-2.1\times 10^{15}$ &$6.5\times 10^5$&$9.1\times 10^{-1}$ \\
$i^s$&   &      & 1.4333 &            &$-1.8\times 10^9$ &                        &
$-9.5\times 10^{14}$ &$6.1\times 10^5$&$1.8\times 10^{-1}$ \\
$i^o$& 3 &      & 1.3591 &$-2.1\times 10^8$ &            &$-1.7\times 10^9$ &
                     &$2.8\times 10^7$&$3.8\times 10^1$ \\
$i^s$&   &      & 1.7829 &$-1.2\times 10^{13}$ &            &$-1.4\times 10^{17}$ &
                     &$1.6\times 10^5$&$1.9\times 10^{-1}$ \\
$i^o$&   &      & 0.5171 &$-4.1\times 10^6$ &            &$-1.9\times 10^6$ &
                     &$8.3\times 10^6$&$1.4\times 10^1$ \\
$i^s$&   &      & 0.6464 &$-3.2\times 10^8$ &            &$-1.8\times 10^{11}$ &
                     &$3.0\times 10^5$&$2.4\times 10^{-1}$ \\
$i^o$& 4 &      & 1.5239 &            &$-1.4\times 10^{13}$&                      &
$-2.7\times 10^{14}$ &$1.6\times 10^7$&$3.3\times 10^{0}$ \\
$i^s$&   &      & 1.9832 &            &$-8.9\times 10^{22}$&                      &
$-2.3\times 10^{34}$ &$1.8\times 10^5$&$1.3\times 10^{-1}$ \\
$i^o$&   &      & 0.8640 &            &$-1.7\times 10^7$   &                      &
$-1.5\times 10^{10}$ &$2.3\times 10^7$&$1.4\times 10^2$ \\
$i^s$&   &      & 1.1063 &            &$-1.2\times 10^6$ &                        &
$-8.9\times 10^{14}$ &$1.8\times 10^5$&$2.3\times 10^{-1}$ \\
$i^o$& 2 & 0.04 & 1.1015 &            &$-3.5\times 10^4$ &                        &
$-4.9\times 10^{14}$ &$2.6\times 10^7$&$2.2\times 10^{1}$ \\
$i^s$&   &      & 1.7632 &            &$-4.1\times 10^{12}$ &                        &
$-8.4\times 10^{19}$ &$4.7\times 10^5$&$4.9\times 10^{-2}$ \\
$i^o$& 3 &      & 1.3605 &$-2.2\times 10^8$ &            &$-1.7\times 10^9$ &
                     &$3.0\times 10^7$&$7.0\times 10^0$ \\
$i^s$&   &      & 2.1661 &$1.2\times 10^{14}$ &            &$5.9\times 10^{18}$ &
                     &$1.3\times 10^5$&$4.5\times 10^{-1}$ \\
$i^o$&   &      & 0.5172 &$-4.4\times 10^6$ &            &$-1.9\times 10^6$ &
                     &$4.1\times 10^6$&$4.2\times 10^0$ \\
$i^s$&   &      & 0.7831 &$-3.0\times 10^8$ &            &$-2.4\times 10^{10}$ &
                     &$1.9\times 10^5$&$6.7\times 10^{-2}$ \\
$i^o$& 4 &      & 1.5188 &            &$-9.8\times 10^{10}$&                      &
$-2.4\times 10^{14}$ &$4.0\times 10^6$&$3.3\times 10^{0}$ \\
$i^s$&   &      & 2.4083 &            &$2.2\times 10^{13}$&                      &
$8.9\times 10^{19}$ &$6.3\times 10^3$&$3.7\times 10^{-2}$ \\
$i^o$&   &      & 0.8642 &            &$-1.9\times 10^7$   &                      &
$-1.5\times 10^{10}$ &$2.2\times 10^7$&$3.9\times 10^1$ \\
$i^s$&   &      & 1.3476 &            &$-1.2\times 10^{10}$ &                        &
$-2.2\times 10^{15}$ &$1.1\times 10^5$&$5.3\times 10^{-2}$ \\
\hline
\end{tabular}
\end{table*}

\newpage

\begin{figure}
\centering
\includegraphics[width=8cm]{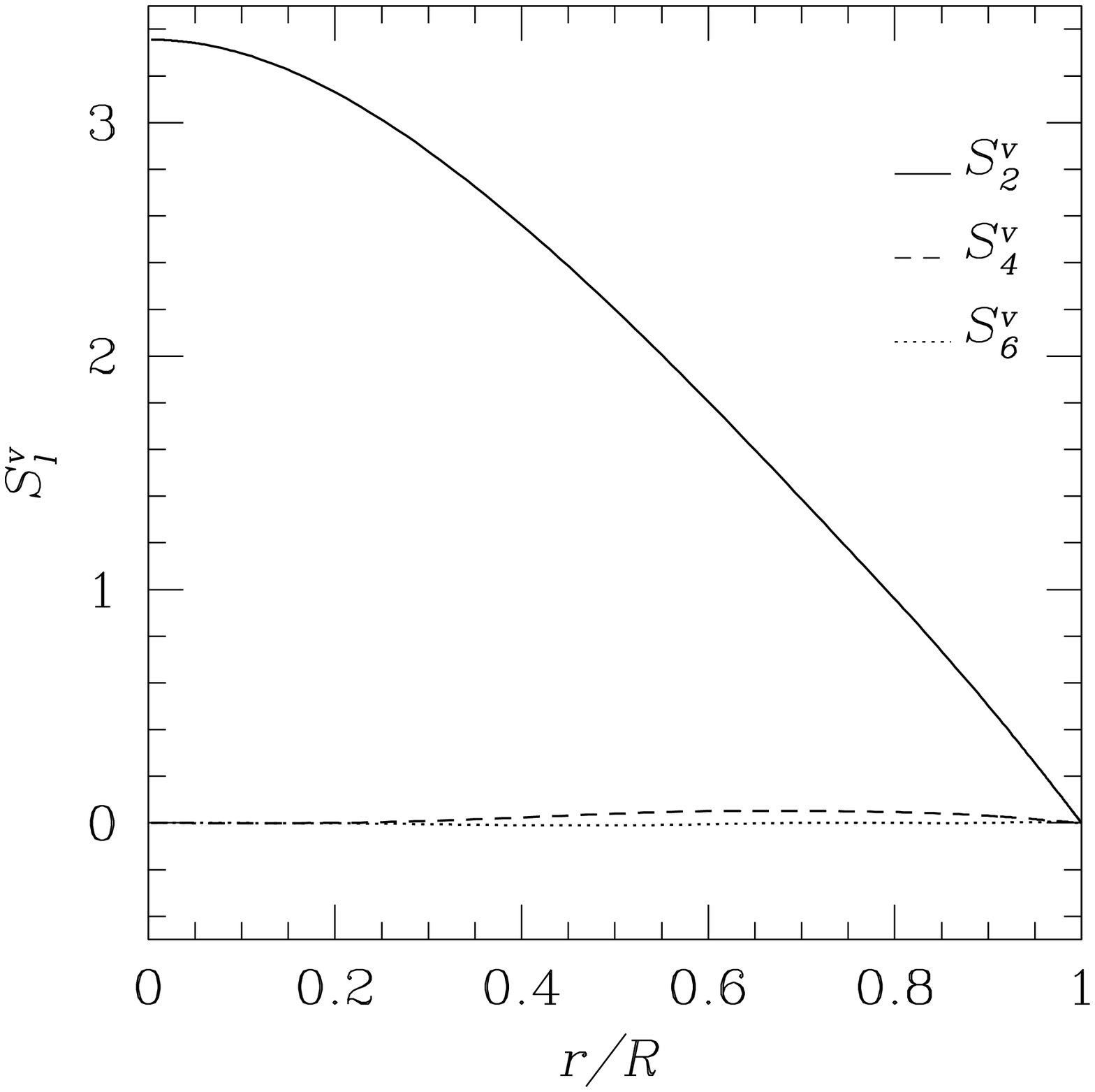}
\includegraphics[width=8cm]{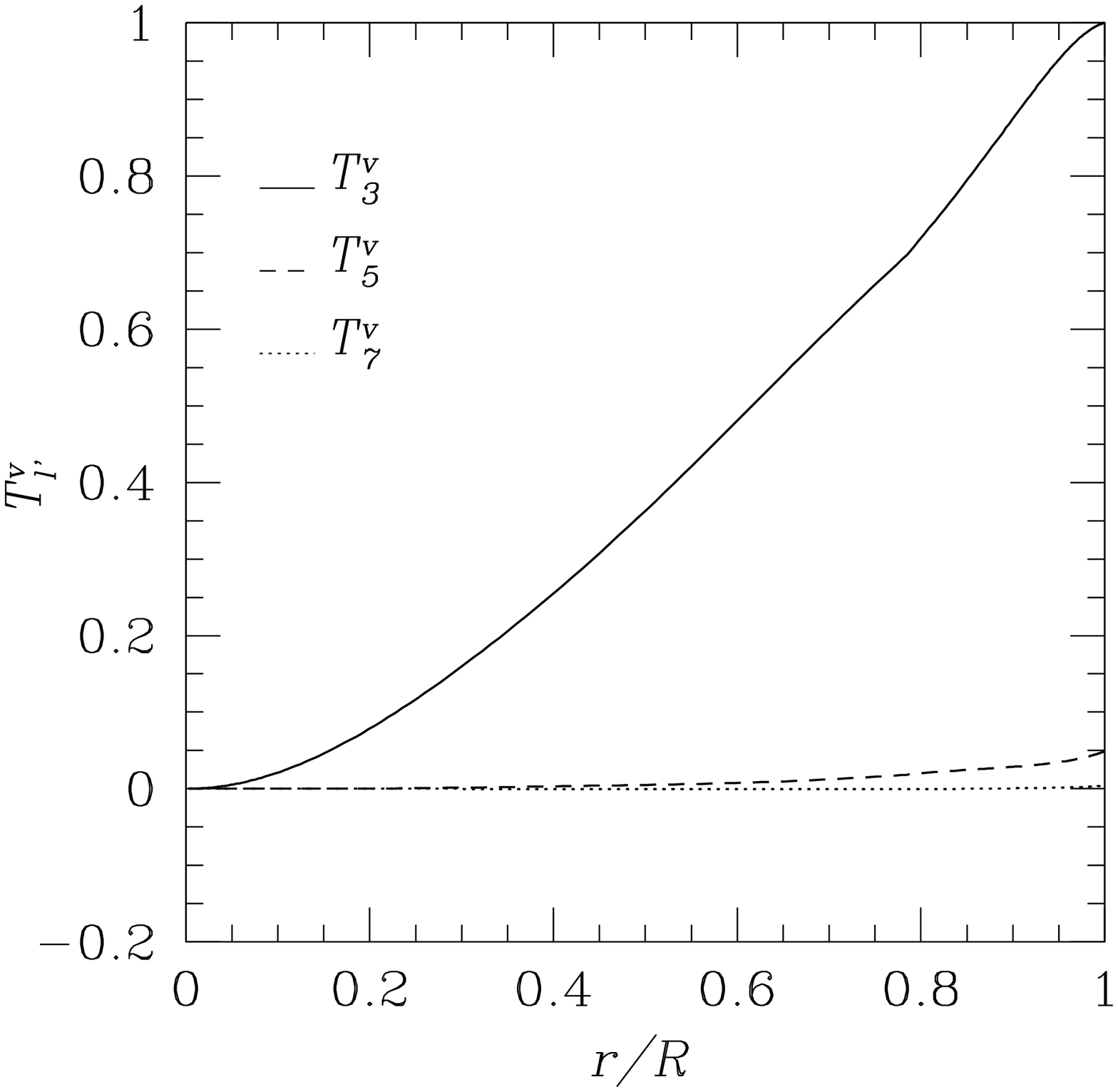}
\caption{First three expansion coefficients $S^v_l$ and $T^v_{l'}$ for the $i^o$-mode with  
$m=2$ and $l_0-|m|=2$ for the case of $\eta=0$, given as a function of $r/R$. 
The mode has $\kappa_0=-0.5686$. The amplitudes are 
normalized by $T^v_{m+1}=1$ at $r=R$. }
\end{figure}

\begin{figure}
\centering
\includegraphics[width=8cm]{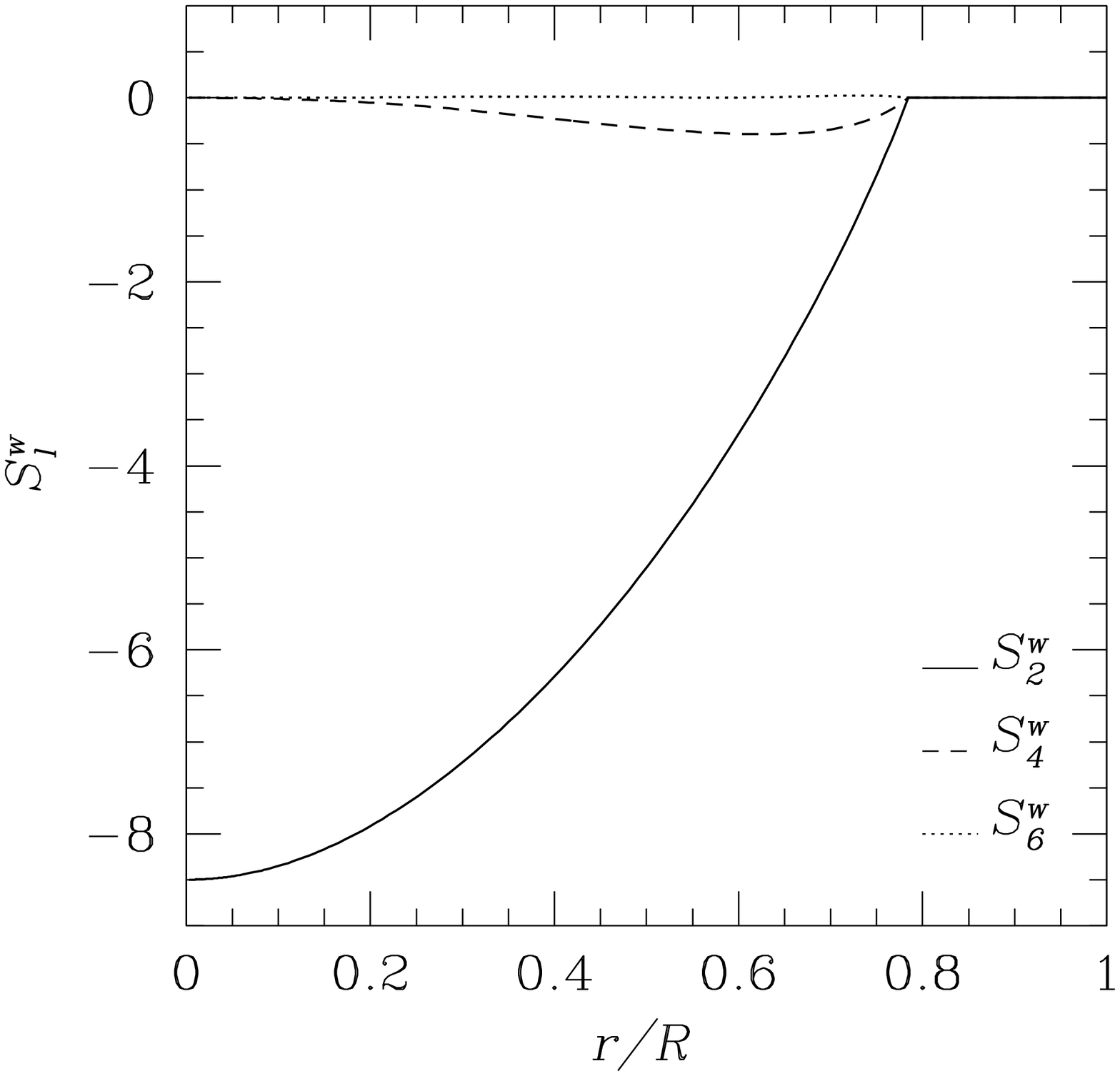}
\includegraphics[width=8cm]{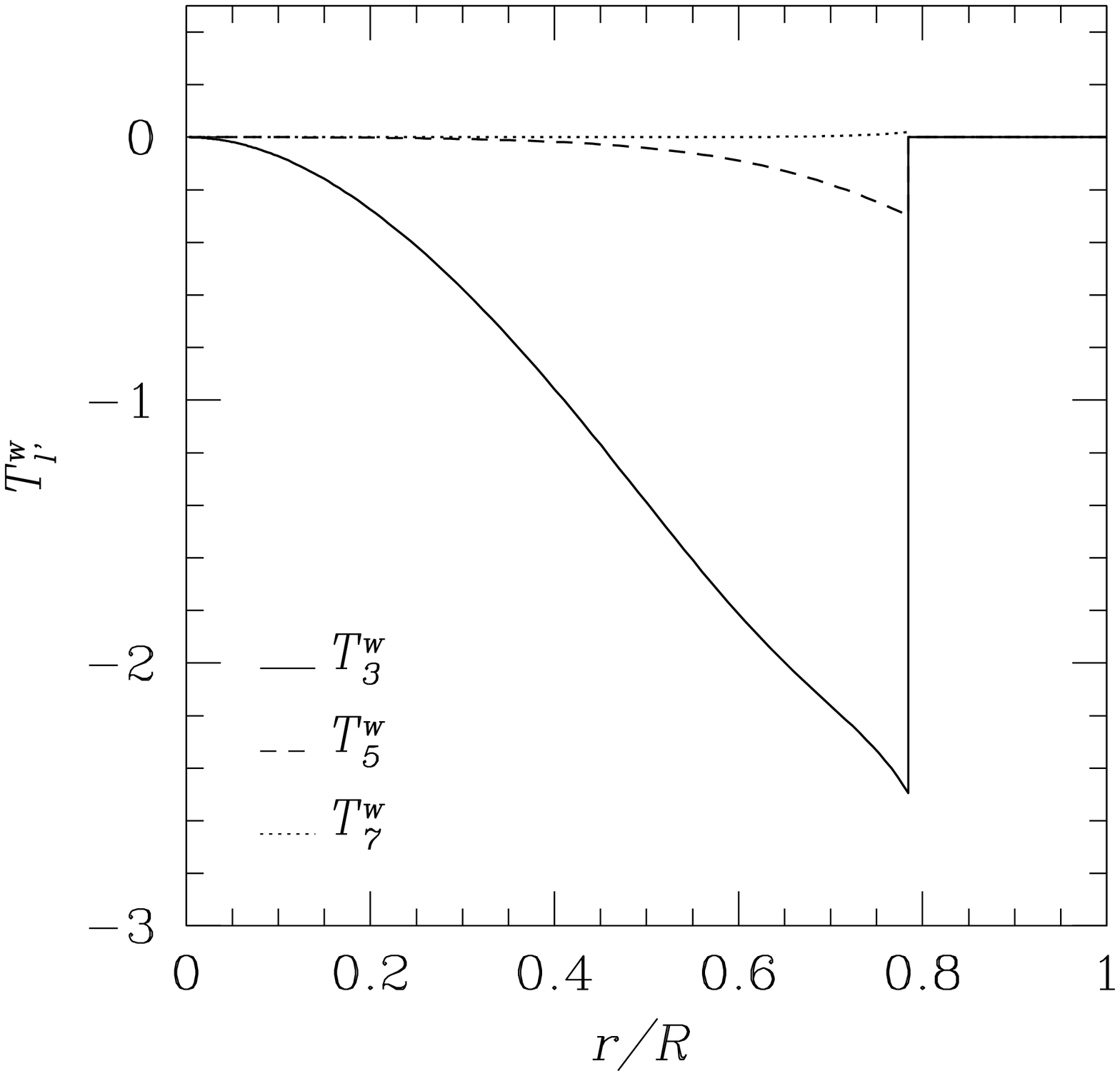}
\caption{Same as Figure 1 but for $S^w_l$ and $T^w_{l'}$.}
\end{figure}

\begin{figure}
\centering
\includegraphics[width=8cm]{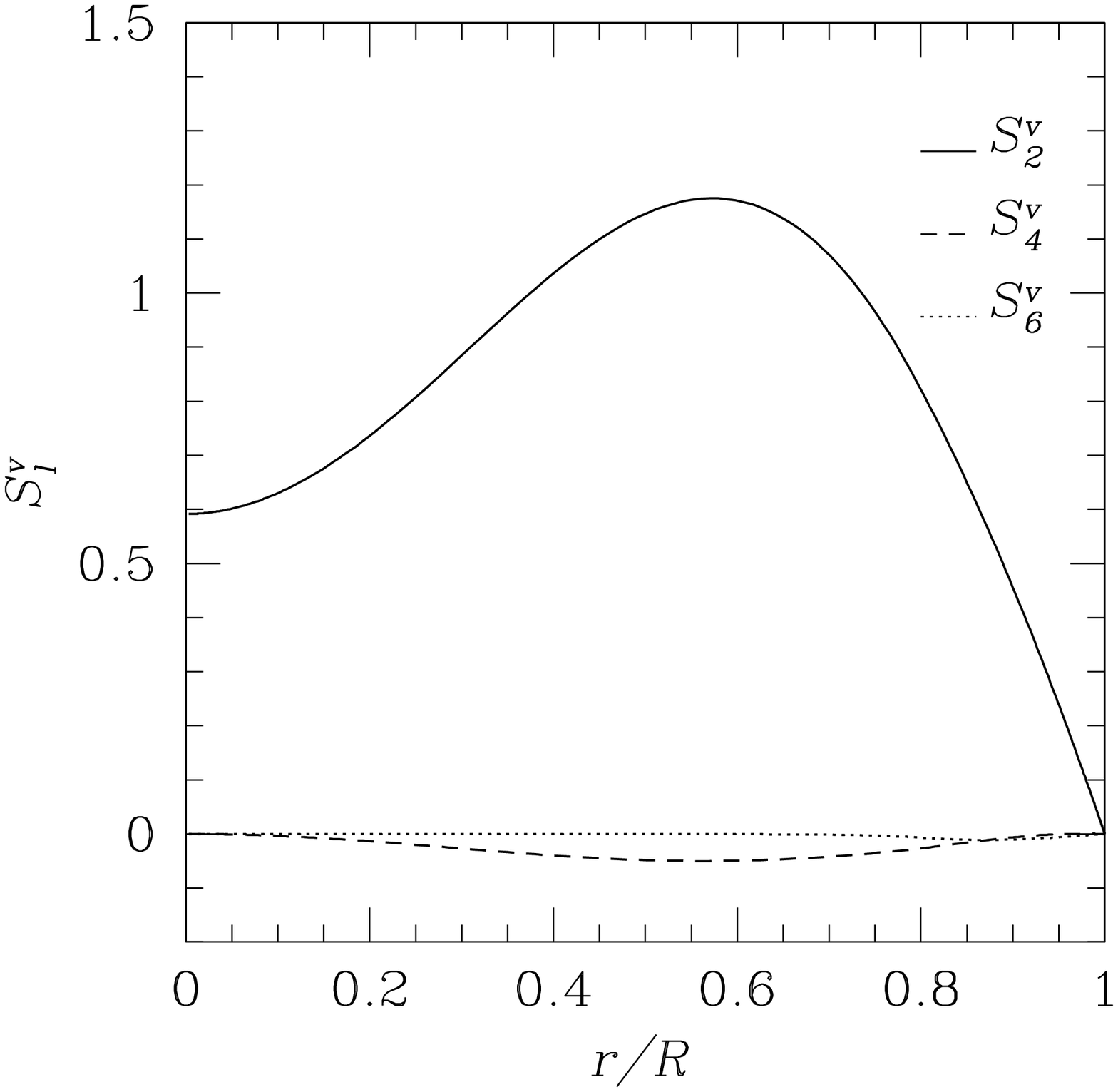}
\includegraphics[width=8cm]{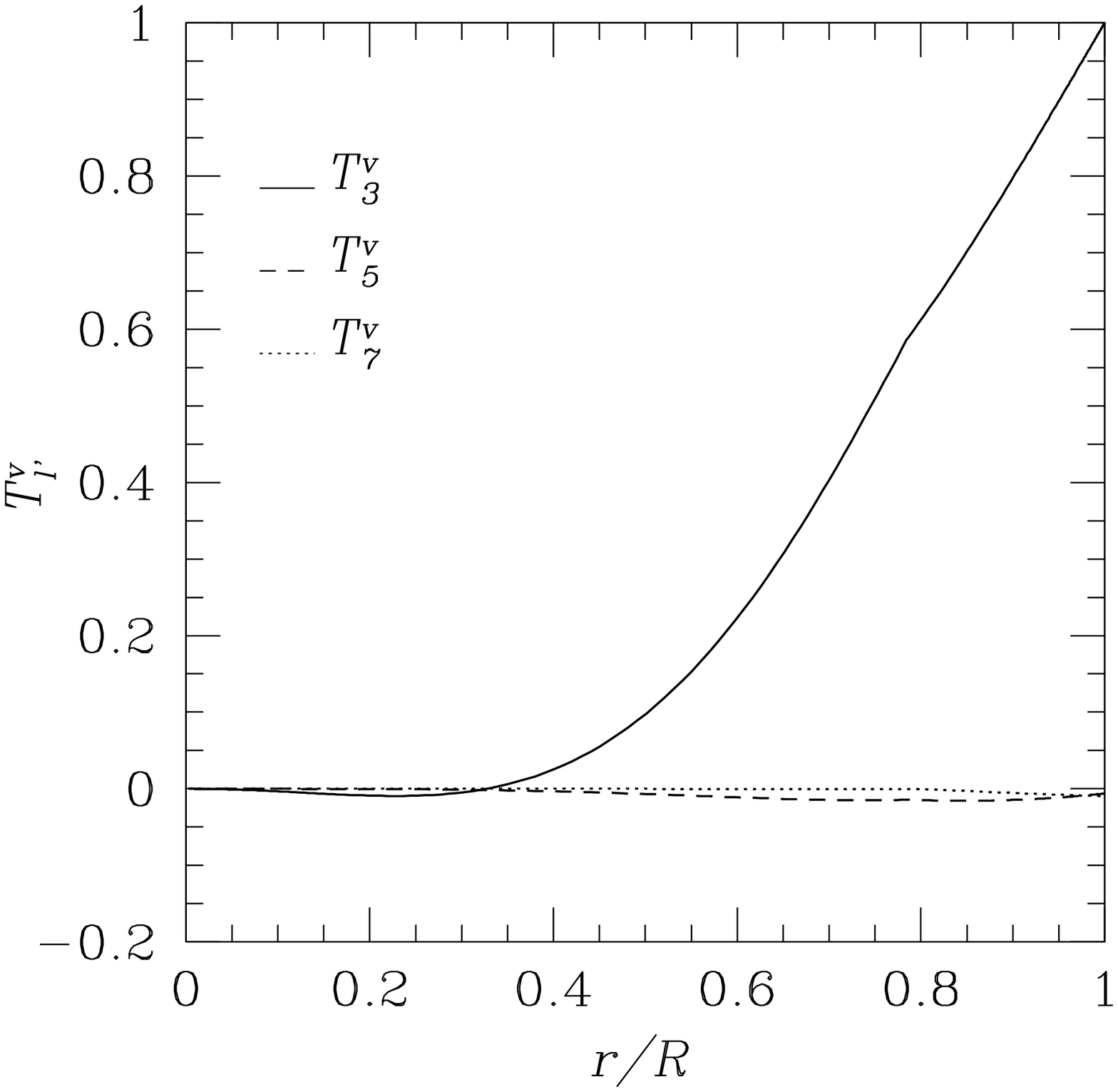}
\caption{First three expansion coefficients $S^v_l$ and $T^v_{l'}$ for the $i^s$-mode with
$m=2$ and $l_0-|m|=2$ for the case of $\eta=0$, given as a function of $r/R$.
The mode has $\kappa_0=-0.5115$. The amplitudes are
normalized by $T^v_{m+1}=1$ at $r=R$. }
\end{figure}

\begin{figure}
\centering
\includegraphics[width=8cm]{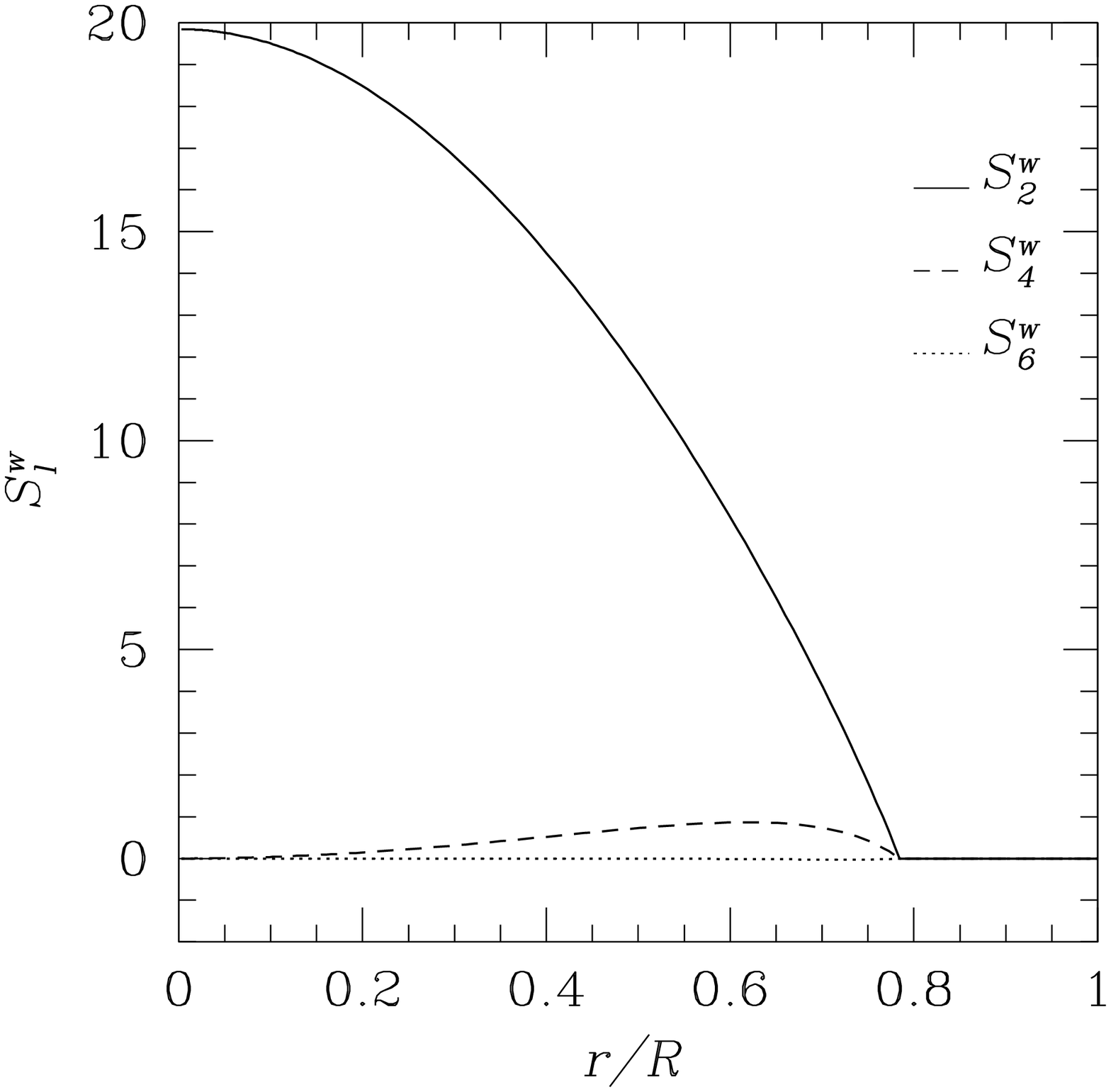}
\includegraphics[width=8cm]{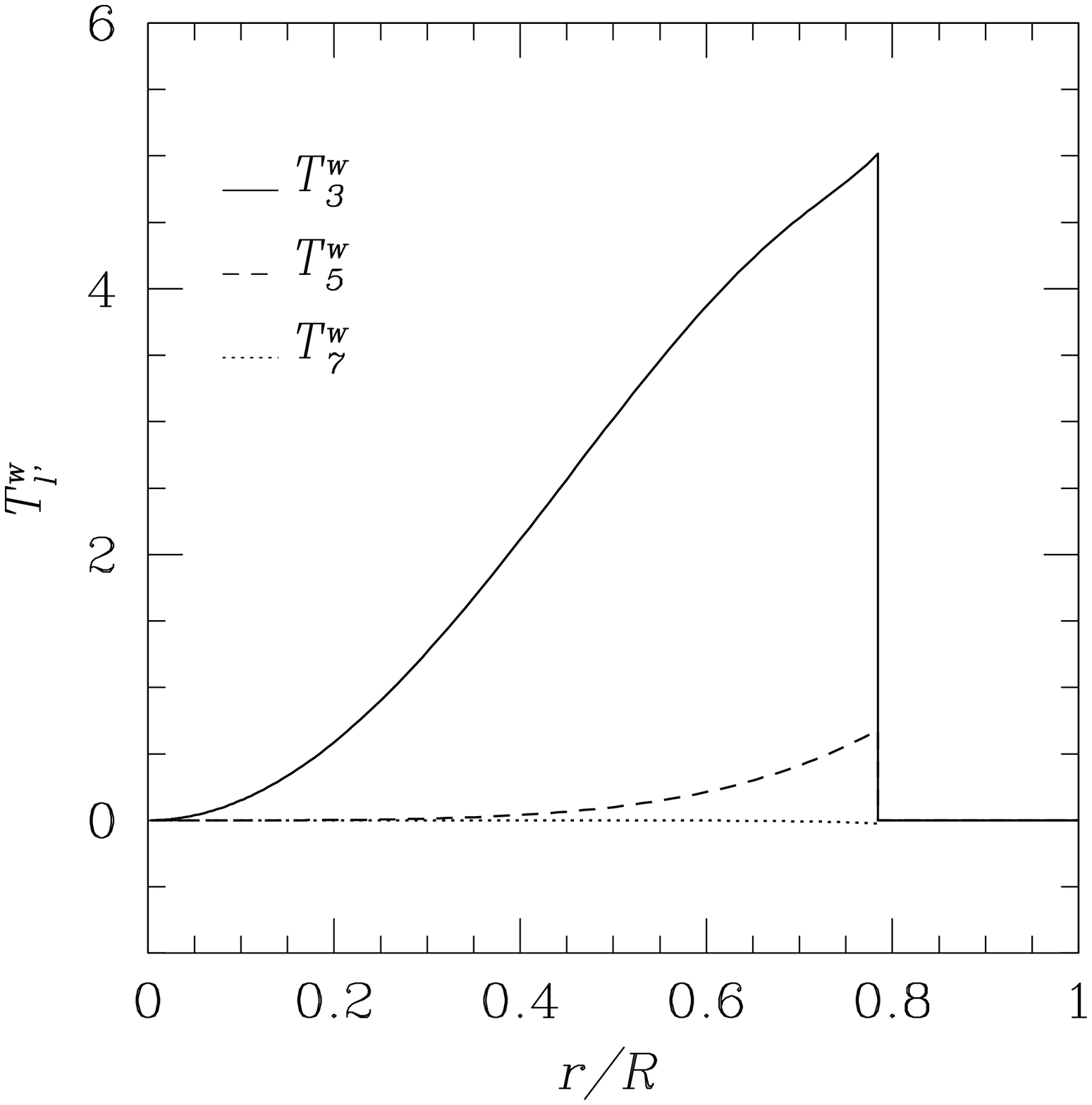}
\caption{Same as Figure 3 but for $S^w_l$ and $T^w_{l'}$.}
\end{figure}

\begin{figure}
\centering
\includegraphics[width=10cm]{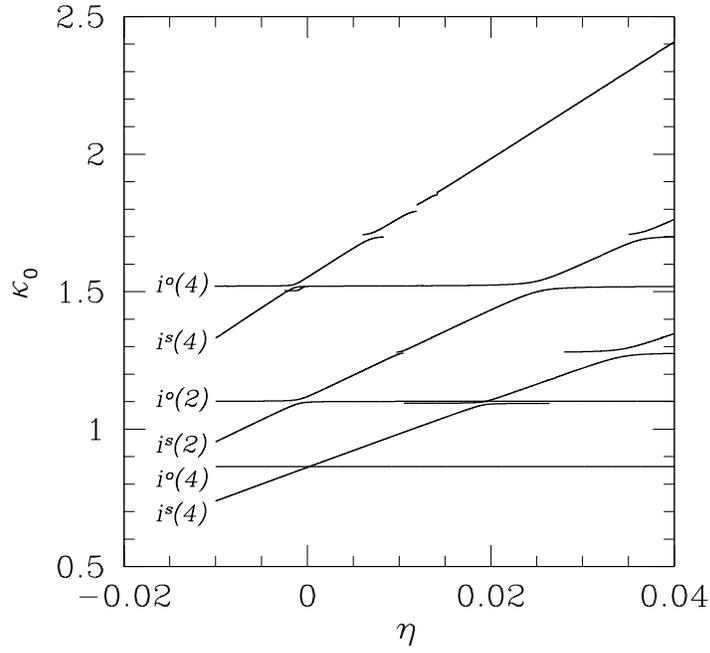}
\caption{$\kappa_0$'s for even parity inertial modes 
having positive $\kappa_0$, given as a function of $\eta$. 
The labels $i^X(k)$ signify the $i^X$-mode 
of $l_0-|m|=k$, where $X$ is ``o'' or ``s''.}
\end{figure}

\begin{figure}
\centering
\includegraphics[width=10cm]{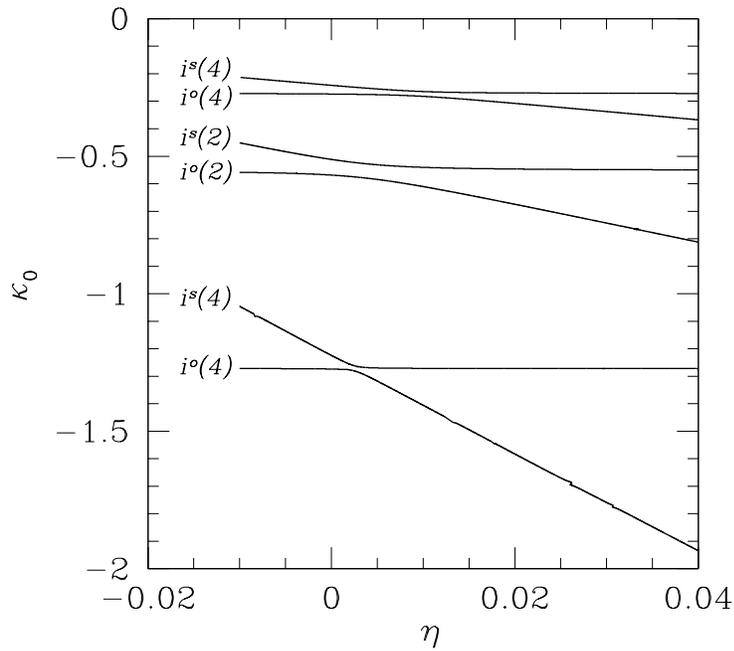}
\caption{Same as Figure 5 but for even parity inertial modes
having negative $\kappa_0$.}
\end{figure}

\begin{figure}
\centering
\includegraphics[width=10cm]{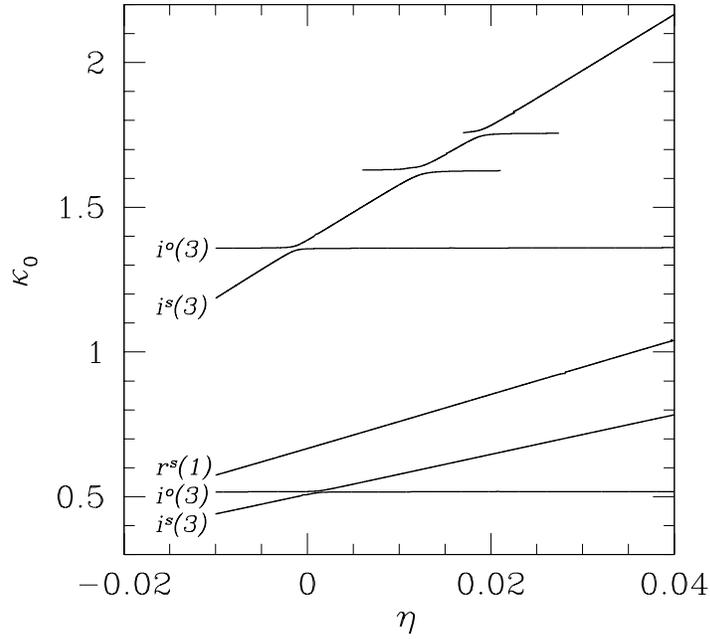}
\caption{$\kappa_0$'s for odd parity inertial modes 
having positive $\kappa_0$, given as a function of $\eta$. 
The labels $i^X(k)$ signify the $i^X$-mode 
of $l_0-|m|=k$, where $X$ is ``o'' or ``s''. Here, $r^s(1)$ indicates
the superfluid $r$-mode, which is the $i^s$-mode of $l_0-|m|=1$. }
\end{figure}

\begin{figure}
\centering
\includegraphics[width=10cm]{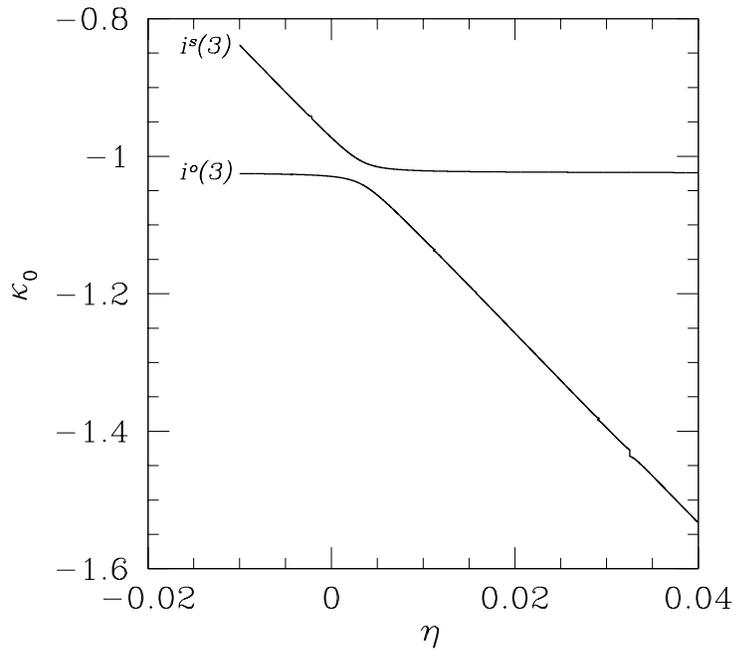}
\caption{Same as Figure 7 but for odd parity inertial modes 
having negative $\kappa_0$.}
\end{figure}

\begin{figure}
\centering
\includegraphics[width=8cm]{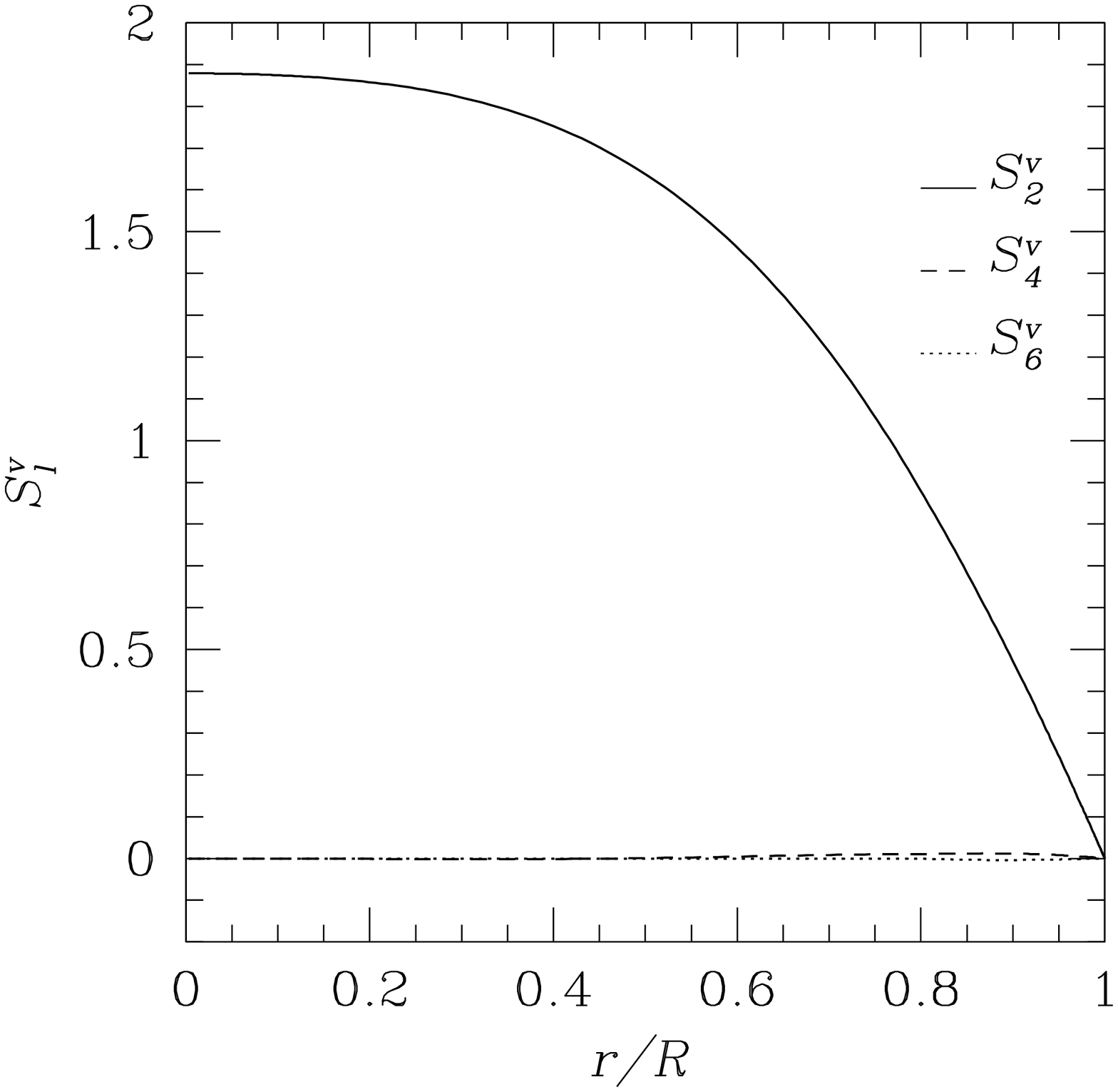}
\includegraphics[width=8cm]{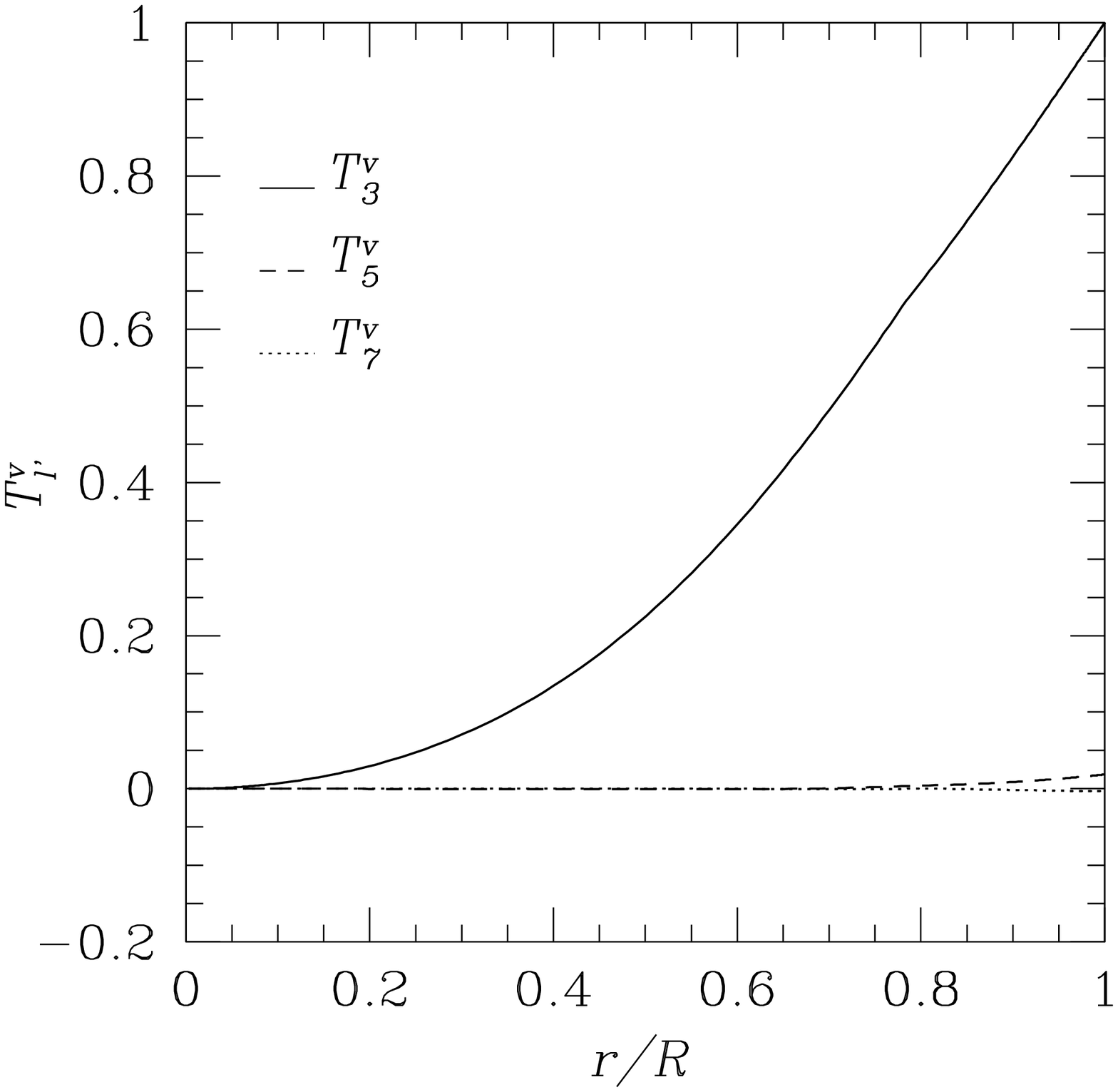}
\caption{First three expansion coefficients $S^v_l$ and $T^v_{l'}$for the $i^o$-mode with
$m=2$ and $l_0-|m|=2$ for the case of $\eta=0.01$, given as a function of $r/R$.
The mode has $\kappa_0=-0.5403$. The amplitudes are
normalized by $T^v_{m+1}=1$ at $r=R$. }
\end{figure}

\begin{figure}
\centering
\includegraphics[width=8cm]{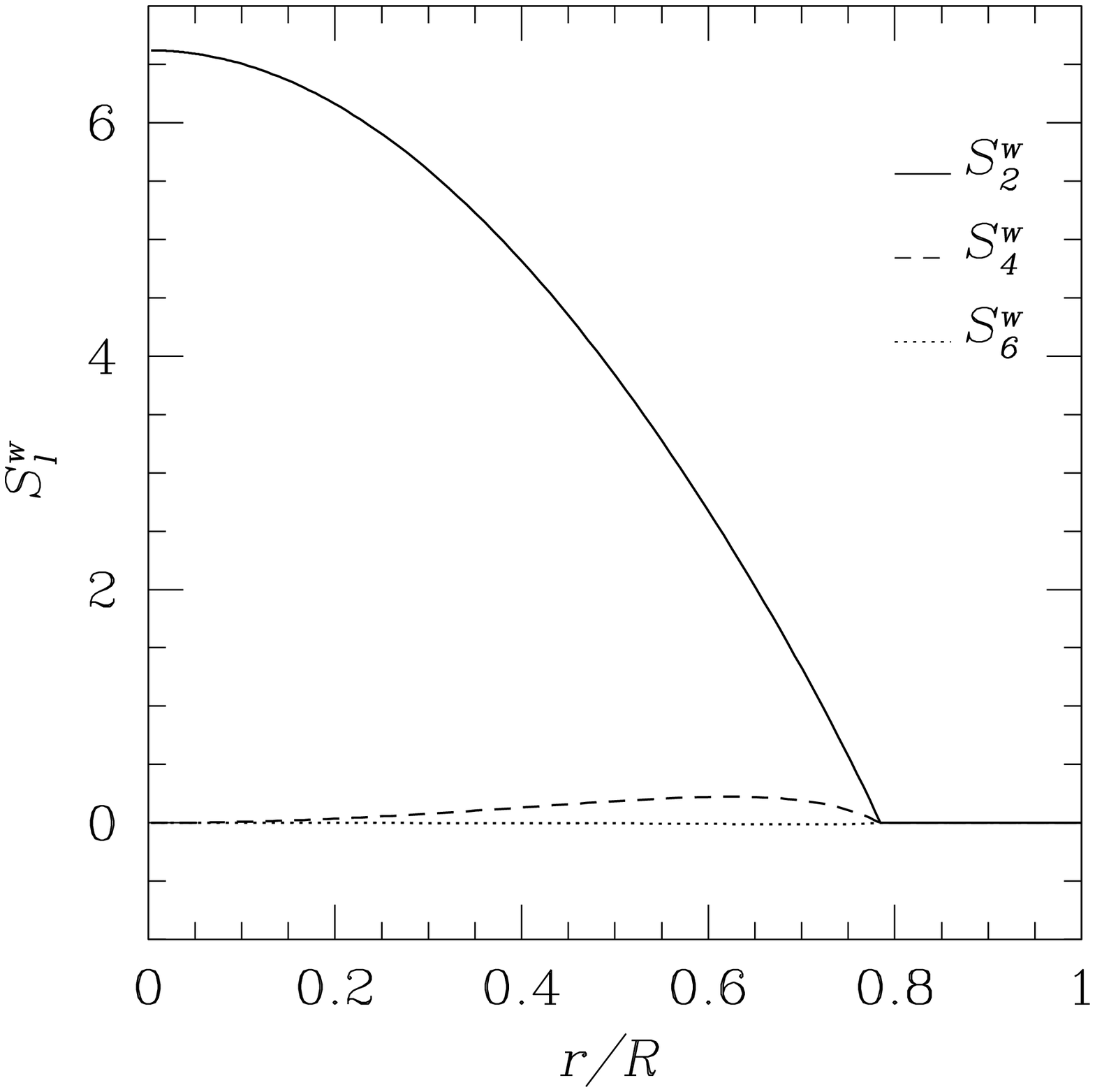}
\includegraphics[width=8cm]{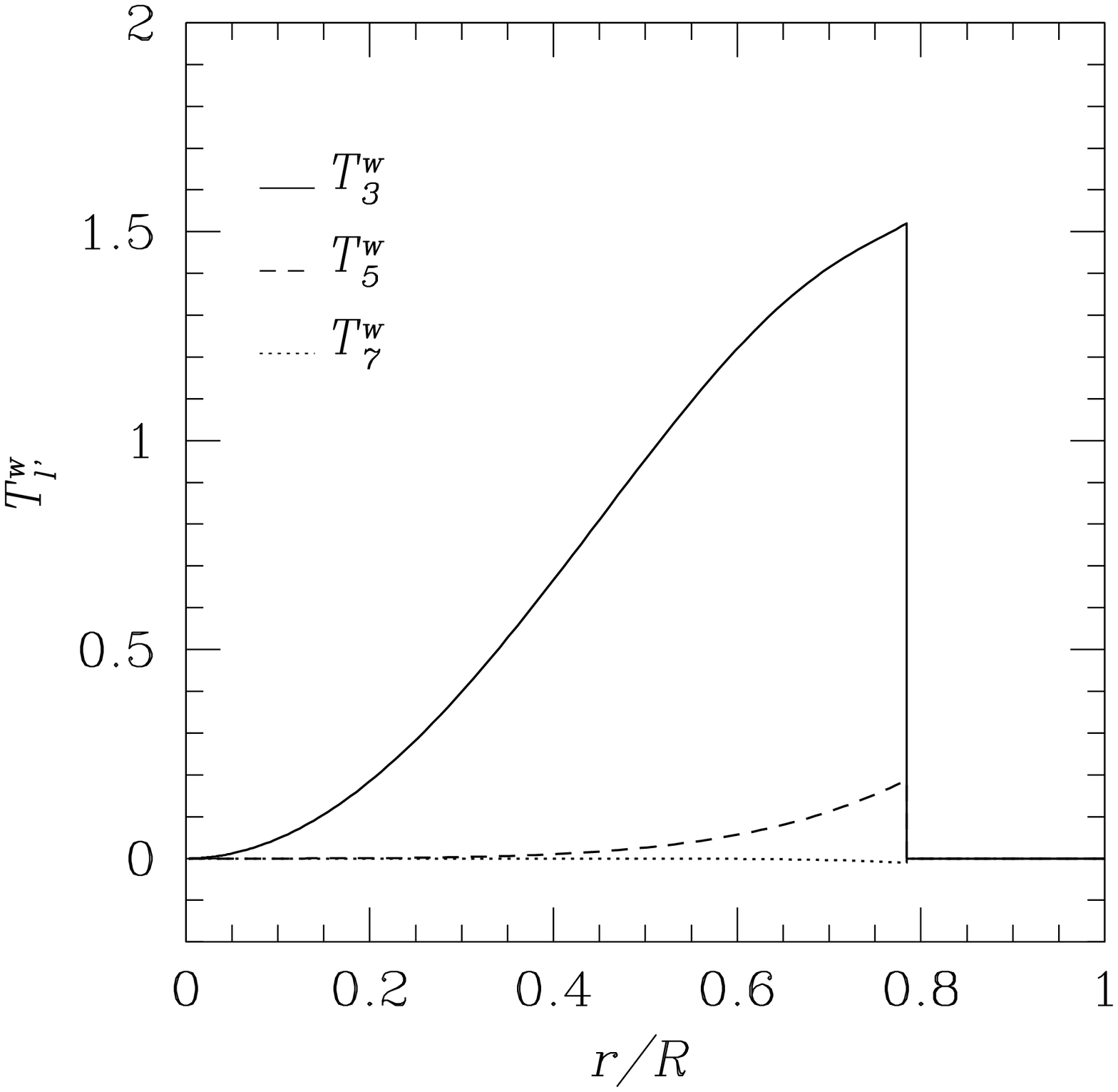}
\caption{Same as Figure 9 but for $S^w_l$ and $T^w_{l'}$.}
\end{figure}

\begin{figure}
\centering
\includegraphics[width=8cm]{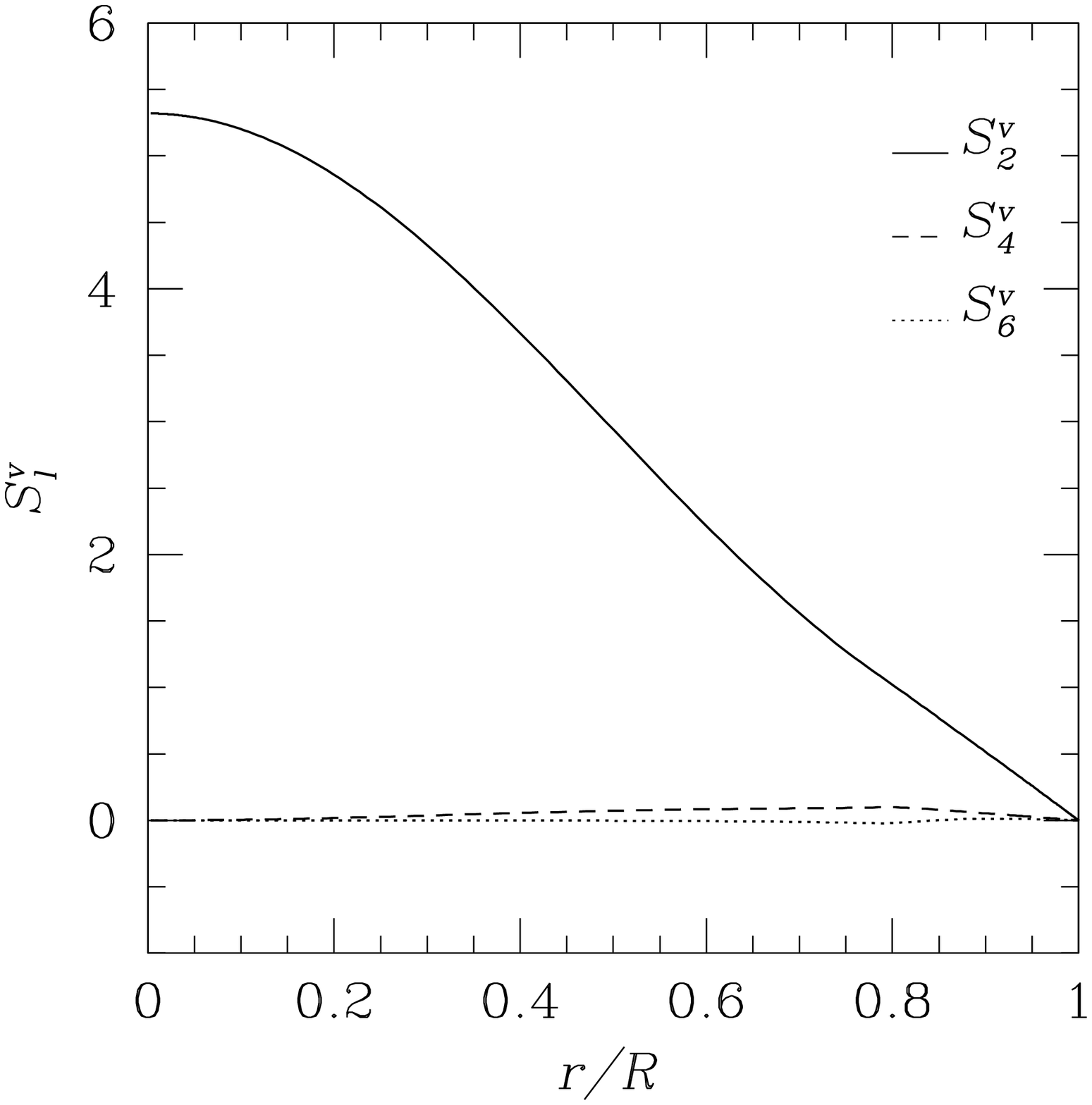}
\includegraphics[width=8cm]{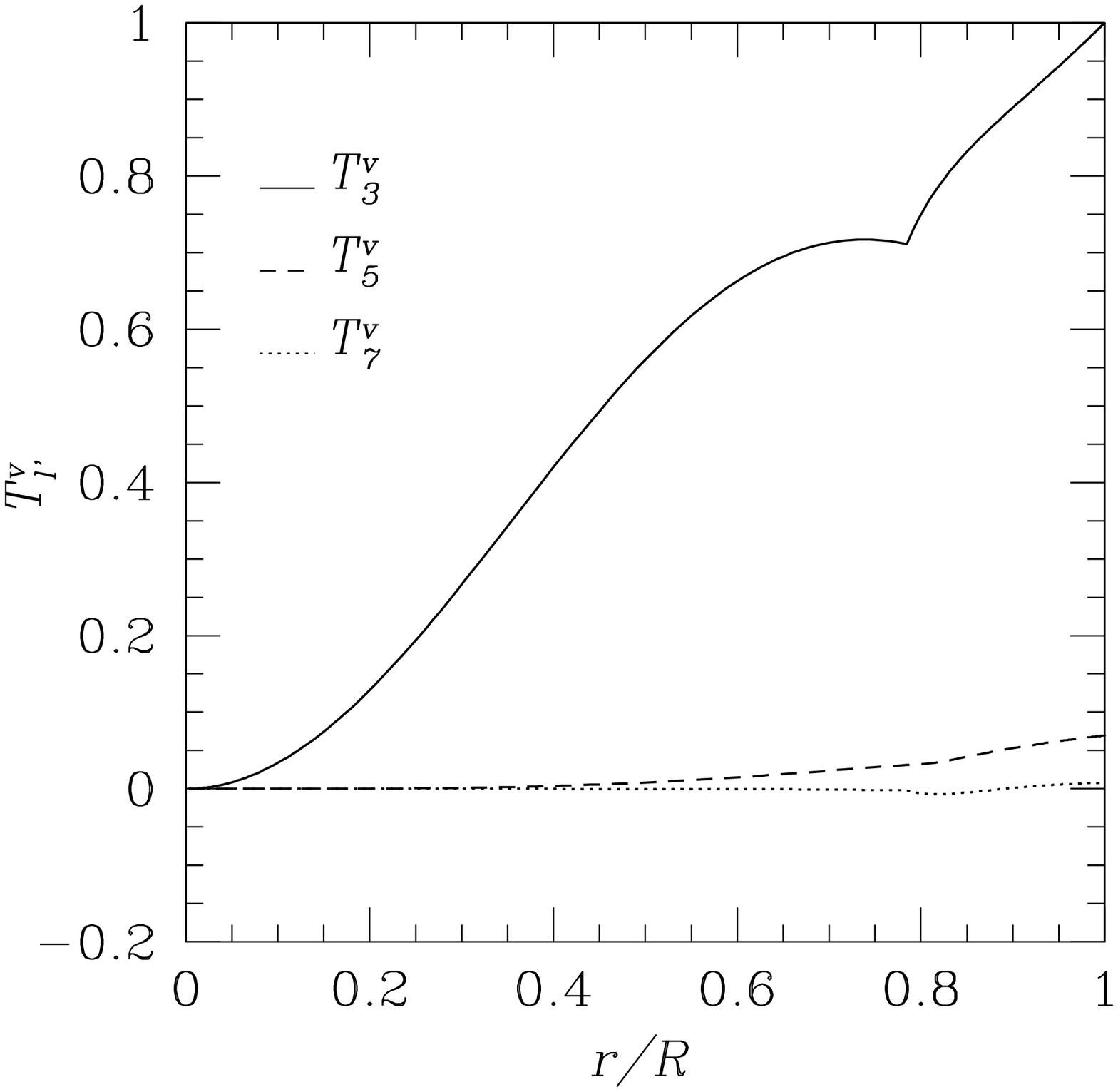}
\caption{First three expansion coefficients $S^v_l$ and $T^v_{l'}$ for the $i^s$-mode with
$m=2$ and $l_0-|m|=2$ for the case of $\eta=0.01$, given as a function of $r/R$.
The mode has $\kappa_0=-0.6103$. The amplitudes are
normalized by $T^v_{m+1}=1$ at $r=R$. }
\end{figure}

\begin{figure}
\centering
\includegraphics[width=8cm]{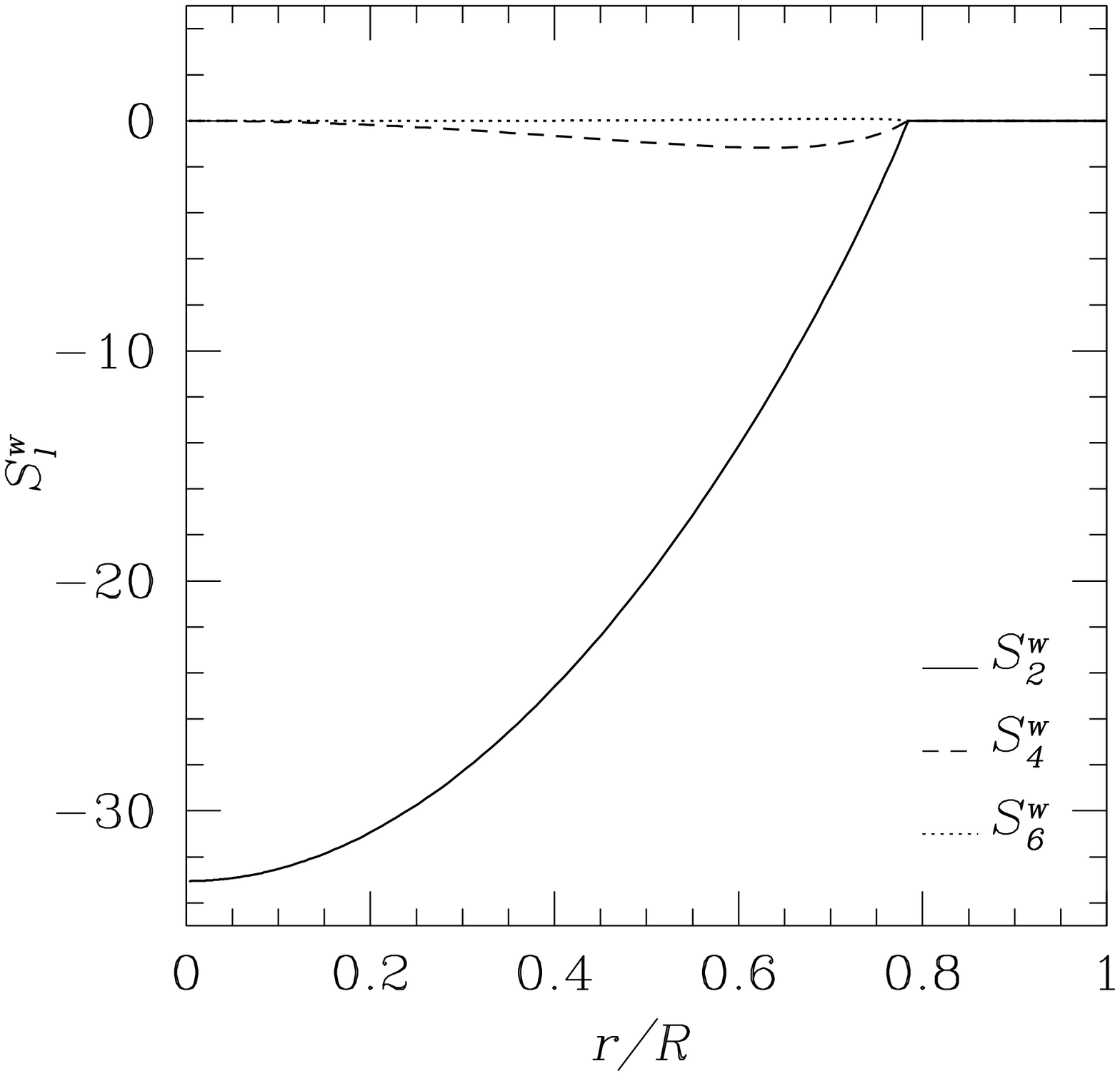}
\includegraphics[width=8cm]{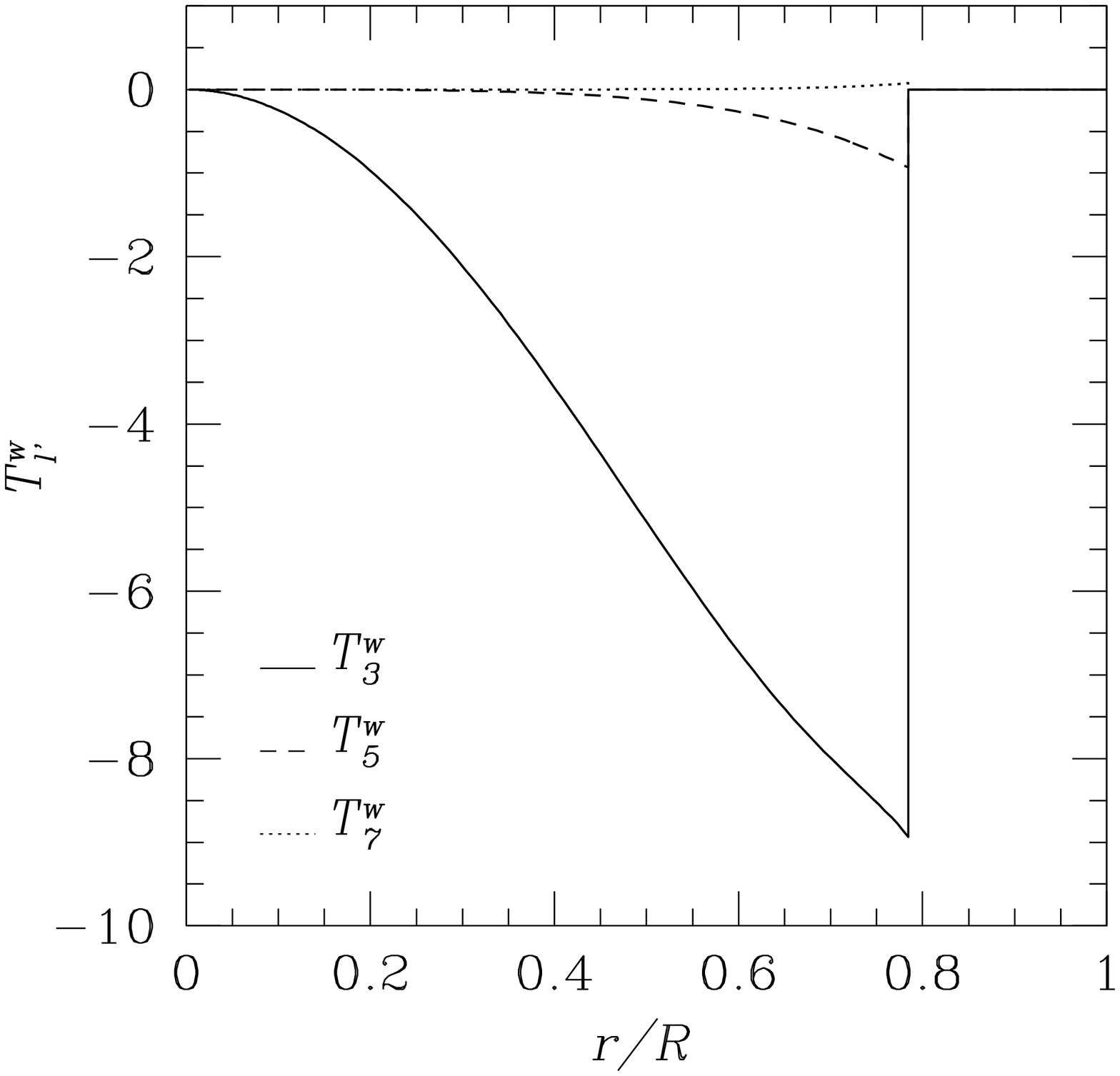}
\caption{Same as Figure 11 but for $S^w_l$ and $T^w_{l'}$.}
\label{fig12}
\end{figure}


\begin{thebibliography}{}

\bibitem[]{} Akmal, A., Pandharipande, V. R., \& Ravenhall, D. G. 1998, Phys. Rev. C, 58, 1804

\bibitem[]{} Alpar, M. A., Langer, S. A., \& Sauls, J. A. 1984, ApJ, 282, 533

\bibitem[]{} Andersson, N. 1998, ApJ, 502, 708

\bibitem[]{} Andersson, N., \& Comer, G. L. 2001, MNRAS, 328, 1129

\bibitem[]{} Andersson, N., Comer, G. L., \& Langlois, D. 2002, Phys. Rev. D, 66, 104002

\bibitem[]{} Andersson, N., \& Kokkotas, K. D. 2001, Int. J. Mod. Phys. D, 10, 381

\bibitem[]{} Andersson, N., Kokkotas, K. D., \& Schutz, B. F. 1999, ApJ, 510, 846

\bibitem[]{} Andersson, N., Kokkotas, K. D., \& Stergioulas, N. 1999, ApJ, 516, 307

\bibitem[]{} Arras, P., Flanagan, E. E., Morsink, S. M., Schenk, A. K., Teukolsky, S. A., 
\& Wasserman, I. 2003, ApJ, in press (astro-ph/0202345). 

\bibitem[]{} Baym, G., \& Pethick, C. 1979, ARA\&A, 17, 415
\bibitem[]{} Bildsten, L. 1998, ApJ, 501, L89

\bibitem[]{} Borumand, M., Joynt, R., \& Klu\'zniak, W. 1996, Phys. Rev. C, 54, 2745

\bibitem[]{} Comer, G. L. 2002, Founds. Phys., 32, 1903 

\bibitem[]{} Comer, G. L., Langlois, D., \& Lin, L. M. 1999, Phys. Rev. D, 60, 104025 

\bibitem[]{} Cutler, L., \& Lindblom, L. 1987, ApJ, 314, 234

\bibitem[]{} Epstein, R. I. 1988, ApJ, 333, 880

\bibitem[]{} Flowers, E., \& Itoh, N. 1979, ApJ, 230, 847

\bibitem[]{} Friedman, J. L., \& Morsink, S. M. 1998, ApJ, 502, 714

\bibitem[]{} Friedman, J. L., \& Schutz, B. F. 1978, ApJ, 222, 281

\bibitem[]{} Lee, U. 1995, A\&A, 303, 515

\bibitem[]{} Lee, U., \& Saio, H. 1986, MNRAS, 221, 365

\bibitem[]{} Lee, U., \& Yoshida, S. 2003, ApJ, 586, 403 

\bibitem[]{} Lindblom, L., \& Ipser, J. R. 1999, Phys. Rev. D, 59, 044009

\bibitem[]{} Lindblom, L., \& Mendell, G. 1994, ApJ, 421, 689

\bibitem[]{} Lindblom, L., \& Mendell, G. 2000, Phys. Rev. D, 61, 104003

\bibitem[]{} Lindblom, L., Owen, B. J., \& Morsink, S. M. 1998, Phys. Rev. Lett, 80, 4843

\bibitem[]{} Lockitch, K. H., \& Friedman, J. L. 1999, ApJ, 521, 764

\bibitem[]{} McDermott, P. N., Van Horn, H. M., \& Hansen, C. J. 1988, ApJ, 325, 725

\bibitem[]{} Mendell, G. 1991a, ApJ, 380, 515

\bibitem[]{} Mendell, G. 1991b, ApJ, 380, 530

\bibitem[]{} Owen, B. J., Lindblom, L., Cutler, C., Schutz, B. F., Vecchio, A, Andersson, N. 
1998, Phys. Rev. D, 58, 084020 

\bibitem[]{} Prix, R., \& Rieutord, M. 2002, A\&A, 393, 949 

\bibitem[]{} Reisenegger, A., \& Goldreich, P. 1992, ApJ, 395, 240

\bibitem[]{} Savonije, G. J., \& Papaloizou, J. C. B. 1983, MNRAS, 203, 581

\bibitem[]{} Sedrakian, A., \& Wasserman, I. 2000, Phys. Rev. D, 63, 024016

\bibitem[]{} Shapiro, S. L., \& Teukolsky, S. A. 1983, Black Holes, White Dwarfs, 
and Neutron Stars (New York: Wiley) 

\bibitem[]{} Thorne, K. 1980, Rev. Mod. Phys., 52, 299

\bibitem[]{} Unno, W., Osaki, Y., Ando, H., Saio, H., \& Shibahashi, H. 1989, 
Nonradial Oscillations of Stars (2d ed.; Tokyo: Univ. Tokyo Press)

\bibitem[]{} Yoshida, S., \& Lee, U. 2000a, ApJ, 529, 997

\bibitem[]{} Yoshida, S., \& Lee, U. 2000b, ApJS, 129, 353

\bibitem[]{} Yoshida, S., \& Lee, U. 2003, Phys. Rev. D, in press (gr-qc/0304073)

\end{thebibliography}
\end{document}